\documentclass{article}

    \PassOptionsToPackage{numbers, compress}{natbib}

\usepackage[main, final]{neurips_2026}

\usepackage[utf8]{inputenc} %
\usepackage[T1]{fontenc}    %
\usepackage{hyperref}       %
\usepackage{url}            %
\usepackage{booktabs}       %
\usepackage{amsfonts}       %
\usepackage{nicefrac}       %
\usepackage{microtype}      %
\usepackage{xcolor}         %
\usepackage{hyperref}
\usepackage{hyperref}
\usepackage{graphicx}
\usepackage{float}
\usepackage{amssymb}\usepackage{amsmath}
\usepackage{algorithm}
\usepackage{algpseudocode}
\usepackage{hyperref}
\usepackage{pifont}
\usepackage{fontawesome5}
\newcommand{\rev}[1]{#1}
\usepackage{wrapfig}

\title{\texorpdfstring{\raisebox{-0.12em}[0pt][0pt]{\includegraphics[height=1em]{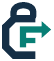}}\hspace{0.3em}}{}\textsc{CipherFlow}: Compiling AI Programs for Efficient Cryptographically Secure Inference}

\author{%
  Hedong Zhang\textsuperscript{\texttt{1}}, Mengxin Zheng\textsuperscript{\texttt{1}}, Qian Lou\textsuperscript{\texttt{1}} \\
  \textsuperscript{\texttt{1}}LoopLeap Lab\\
  \texttt{\{hedong.zhang,mengxin.zheng,qian.lou\}@ucf.edu}
}

\begin{document}

\maketitle

\begin{center}
    \vspace{-1.5em}
    \small
    \href{https://github.com/loopleap-lab/CipherFlow}{\faGithub\enspace Project Page:https://github.com/loopleap-lab/CipherFlow}
\end{center}
\vspace{0.8em}

\begin{abstract}
Cryptographically secure inference faces two practical challenges: making secure programming accessible to AI developers and data scientists, and achieving efficient execution across deployment environments. Developers must compose cryptographic primitives correctly, while performance depends on hardware, networks, and evolving ciphertext states. We present \textsc{CipherFlow}, a compiler framework that takes a high-level AI computation graph, available cryptographic primitives, and a target hardware and network configuration to generate an executable, latency-optimized secure inference program. \textsc{CipherFlow} automates cryptographic execution decisions through three techniques: \textit{MicroBench} profiles deployment-specific primitive costs; state-expanded planning jointly optimizes homomorphic encryption (HE) and secure multi-party computation (MPC) placement, conversions, and ciphertext maintenance under cryptographic-state constraints; and \textit{BlockFold} summarizes branched regions to scale planning to Transformer graphs. Evaluation using measured HE/MPC backend costs across CPU/GPU and network settings shows up to 16.5$\times$ speedup on BERT-base and over 22$\times$ on ViT compared with static hybrid placement baselines. \textsc{CipherFlow} separates model programming from cryptographic execution decisions, enabling developers to express AI computations while the compiler constructs efficient secure programs.
\end{abstract}

\begin{figure}[H]
    \centering
    \includegraphics[width=0.85\linewidth]{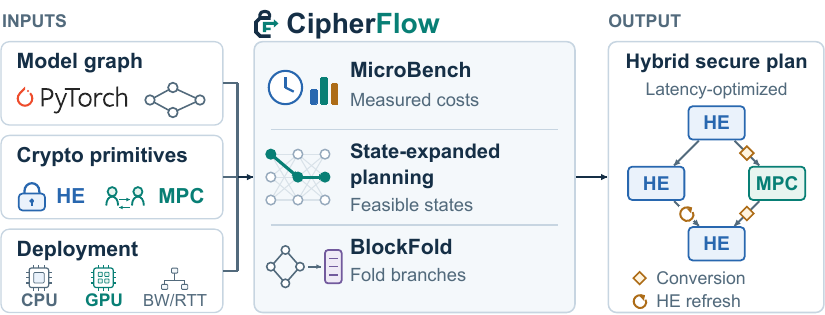}
    \caption{\textbf{CipherFlow at a glance.} Model graphs, cryptographic primitives, and hardware/network profiles inform a latency-optimized HE--MPC plan with conversions and HE maintenance.}
    \label{fig:cipherflow-overview}
\end{figure}

\section{Introduction}
\label{sec:intro}
AI developers and data scientists increasingly deploy models in privacy-sensitive domains such as healthcare, finance, and biometric authentication, where plaintext cloud inference can expose user inputs and model intellectual property~\cite{zhang2025zkvc,shamshad2023transformers,cheng2022personal,sarker2024transformer,rathee2020cryptflow2,huang2022cheetah}. Cryptographically secure inference protects these assets by evaluating neural networks with homomorphic encryption (HE)~\cite{gentry2009fully} and secure multi-party computation (MPC)~\cite{yao1986generate}, without exposing inputs, model weights, or intermediate activations beyond the prescribed output~\cite{zhang2025attestllm,zhang2024secure,xue2025securing,lou2021safenet,xu2021privacy,kei2025shaft}. Turning this capability into a practical way to deploy AI requires addressing two coupled challenges: \textbf{programming usability} and \textbf{execution efficiency}.

\textbf{Challenge 1: Programming usability.}
AI programmers express models through tensor operators and data dependencies; a secure implementation must additionally specify how those computations are realized through cryptographic protocols. The developer must choose HE or MPC for each operator, insert conversions between representations, and maintain compatible ciphertext states along sequential paths and at branch joins~\cite{lou2020autoprivacy,lou2021hemet}. These decisions require cryptographic and systems expertise beyond model development. For example, assigning a matrix multiplication to HE is insufficient if its input ciphertext lacks the required level budget, and a residual addition cannot execute in HE until its inputs have compatible states. Prior secure-computation compilers and mixed-protocol frameworks establish the value of translating high-level programs into cryptographic execution~\cite{chandran2019ezpc,boemer2020mp2ml}. The programming challenge is therefore to let developers specify the AI computation while a compiler constructs a complete, executable secure program, including the conversions and maintenance actions needed by its backend primitives.

\begin{figure}[ht]
    \centering
    \includegraphics[width=1.0\linewidth]{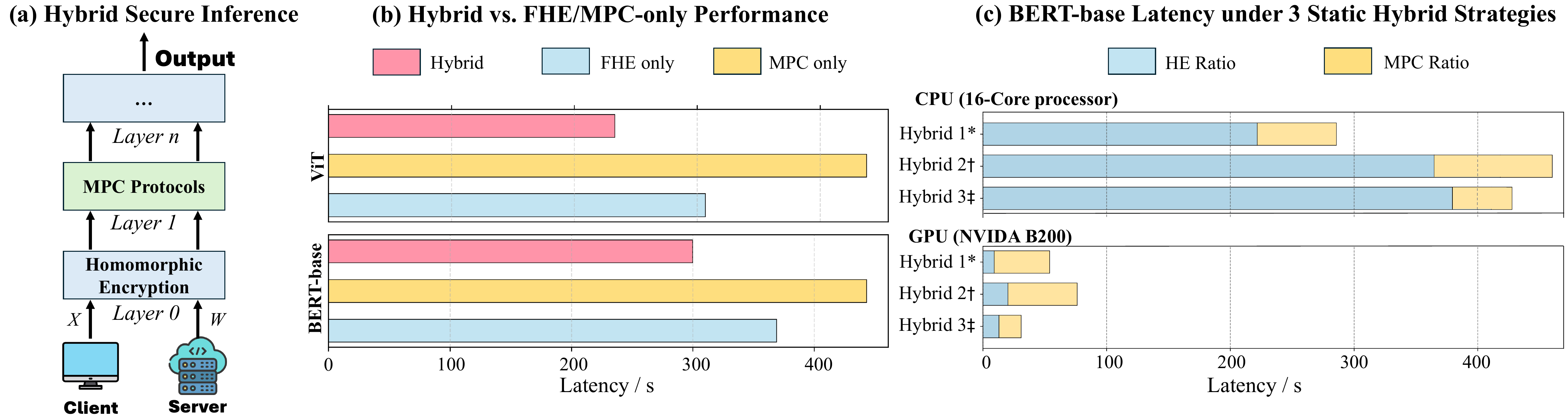}
    \vspace{-1.5 em}
    \caption{\textbf{(a)} Hybrid secure inference executes different operators in HE or MPC with explicit conversions.  \textbf{(b)} Hybrid execution can outperform FHE-only and MPC-only across different model architectures under the same CPU (16-core) and network setting \(200\,\mathrm{Mbps}\), \(4\,\mathrm{ms}\).  \textbf{(c)} BERT-base latency under three static hybrid strategies across different hardware settings. ($\ast$ Linear$\in$ HE, Nonlinear$\in$ MPC~\cite{xu2025breaking}; $\dag$ Attention$\in$ HE, FFN$\in$ MPC; $\ddag$ Attention$\in$ MPC, FFN$\in$ HE)}
    \label{fig:intro}
\end{figure}

\textbf{Challenge 2: Execution efficiency.}
An executable secure program must also adapt to the deployment on which it runs. HE can efficiently process linear operations such as matrix multiplications and convolutions with little interaction~\cite{huang2022cheetah,juvekar2018gazelle,wahyudi2026efficient}, whereas nonlinear functions such as ReLU and Softmax can require expensive approximations or ciphertext maintenance~\cite{zhang2024secure,lou2019she,ebel2025orion,moon2024thor,zhang2023hebridge,yudha2024boostcom,xue2025sok}. MPC offers alternative protocols for these functions but incurs communication and synchronization costs~\cite{riazi2018chameleon,knott2021crypten}. This complementarity motivates hybrid HE--MPC execution~\cite{xue2025sok,lou2020autoprivacy,zhang2025cipherprune}. Figure~\ref{fig:intro}(a,b) illustrates how explicit conversions connect the two domains and how a hybrid strategy can outperform either domain alone. Representative hybrid systems use manually designed placement strategies~\cite{zheng2023primer,hao2022iron,xu2025breaking,lu2023bumblebee}, but a fixed assignment by operator type cannot capture changing compute and communication costs.

Execution efficiency depends on both deployment costs and cryptographic state. GPU acceleration changes the cost of HE relative to MPC~\cite{zhang2024secure,yang2024phantom,jiang2022matcha,zheng2023priml,deng2024trinity,zheng2024ofhe,han2022coxhe,li2026reliafhe,zheng2022cryptolight}, while bandwidth and round-trip time change the cost of interactive protocols and cross-domain conversions~\cite{huang2022cheetah,pang2024bolt,fang2022costco,watson2022piranha}. Figure~\ref{fig:intro}(c) shows that the best static strategy for BERT-base changes across CPU and GPU deployments. Moreover, HE operations consume a finite ciphertext level budget~\cite{fan2012somewhat,cheon2017homomorphic}; continuing in HE may require bootstrapping~\cite{gentry2009fully} or a conversion round trip. At residual and attention joins, incoming states must also be compatible. Consequently, a locally inexpensive operator choice can force costly maintenance or conversions downstream, and enumerating joint states across branches can make planning intractable for Transformer graphs~\cite{vaswani2017attention}. Efficient compilation must optimize these coupled decisions while enforcing execution feasibility.

\textbf{CipherFlow: from AI programs to secure execution.}
We present \textsc{CipherFlow} (Figure~\ref{fig:cipherflow-overview}), a compiler framework that takes a high-level AI computation graph, available cryptographic primitives, and a target hardware and network configuration, and generates a deployment-specialized, latency-optimized secure inference program. It addresses programming usability by separating model computation from cryptographic execution decisions: the compiler selects HE/MPC operators and materializes conversions, state alignment, and maintenance actions using the supported backends. It addresses execution efficiency by jointly planning these actions under measured deployment costs and cryptographic-state constraints. The resulting program composes the backend primitives under their security assumptions; the compiler automates their selection and execution planning.

\textsc{CipherFlow} combines three techniques. \textit{MicroBench} profiles HE/MPC operators, conversions, and bootstrapping on the target deployment. It benchmarks representative secure operator patterns and reuses measurements across repeated model blocks, keeping profiling costs manageable. \textit{State-expanded planning} represents each operator together with its execution domain and remaining HE level, and jointly optimizes placement, conversion, and maintenance decisions. This turns feasibility-aware planning on sequential regions into shortest-path search. \textit{BlockFold} extends planning to branched graphs by summarizing residual and attention regions as boundary-state transition tables. These summaries capture feasible, minimum-cost transformations through each region and support their composition without enumerating all branch-internal state combinations across the full graph. Together, the three techniques connect a high-level programming interface to executable secure computation tailored to its deployment.

We make four contributions:
\begin{enumerate}
    \item \textbf{A compiler framework for usable and efficient secure AI inference.} We formulate compilation from AI computation graphs, cryptographic primitives, and deployment configurations to executable hybrid secure programs, jointly capturing operator placement, conversions, ciphertext maintenance, and cryptographic-state feasibility.
    \item \textbf{Deployment-specific cost profiling and state-aware planning.} We design \textit{MicroBench} and a state-expanded graph formulation that jointly optimize HE/MPC execution and auxiliary actions under measured hardware and network costs, reducing sequential planning to shortest-path search.
    \item \textbf{Scalable planning for branched neural graphs.} We introduce \textit{BlockFold}, which summarizes branched regions using boundary-state transition tables and composes them into a global plan while respecting compatibility at joins, enabling optimization on Transformer-scale graphs.
    \item \textbf{Implementation and evaluation with real cryptographic backends.} We implement \textsc{CipherFlow} on HE/MPC backends and evaluate plans using measured operator latency, communication, and maintenance costs across CPU/GPU and network settings. On BERT-base and ViT, the evaluation shows up to 16.5$\times$ and over 22$\times$ speedup, respectively, compared with static hybrid placement baselines~\cite{xu2025breaking,lu2023bumblebee,pang2024bolt}.
\end{enumerate}

\section{Preliminaries \& Problem Formulation}
\label{sec:prelim}
\subsection{Preliminaries}
\label{sec:preliminaries}
We summarize the primitives and cost quantities used by the compiler.

\noindent\textbf{Secure Multi-Party Computation.}
We use semi-honest two-party computation (2PC) over additive secret shares~\cite{chandran2019ezpc,yao1982protocols}: a value $x$ is split as $[x]_0+[x]_1\equiv x\pmod{2^\ell}$. Linear operations are evaluated locally, while nonlinear operations invoke interactive sub-protocols~\cite{knott2021crypten,rathee2021sirnn}. Thus, MPC latency depends on local compute, communication volume $b$, and round count $r$ under bandwidth $\mathrm{BW}$ and round-trip time ($\mathrm{RTT})$.

\noindent\textbf{Homomorphic Encryption.}
We use a leveled CKKS-style scheme~\cite{zhang2024secure,cheon2017homomorphic}. Each ciphertext carries a finite level budget $L$; executing operator $v_i$ in HE consumes $\delta_i$ levels, and if the remaining level $l<\delta_i$, a costly bootstrap is required to refresh the ciphertext~\cite{gentry2009fully}. HE additions at residual or attention joins further require compatible ciphertext levels.

\noindent\textbf{HE-MPC Conversion.}
Hybrid execution moves intermediate values across the two domains via HE$\rightarrow$MPC and MPC$\rightarrow$HE. Both carry measurable compute and communication cost. \rev{An HE$\rightarrow$MPC$\rightarrow$HE round trip restores a fresh HE state and can therefore substitute for a bootstrap, but it is not free: it pays two conversions, whose communication-dominated cost trades off against the compute-bound cost of a direct bootstrap depending on the deployment (Section~\ref{sec:state-expanded}).}%

\paragraph{Compilers and cost models for secure computation.}
Prior compilers and mixed-protocol frameworks, including EzPC and MP2ML, translate high-level programs into secure-computation protocols~\cite{chandran2019ezpc,boemer2020mp2ml}, while HE compilers such as nGraph-HE2 optimize neural-network inference within a single HE backend~\cite{kumar2026fhe,zheng2023cofhe,boemer2019ngraph}. Cost-modeling systems such as COSTCO show that secure-computation performance depends strongly on operator shape, hardware, and network conditions~\cite{fang2022costco}. Our compiler is complementary but targets a different optimization problem: it jointly plans HE/MPC placement, bidirectional conversions, and HE level maintenance over branched DNN graphs.

\paragraph{Private neural-network and Transformer inference optimizations.}
Other work improves private inference by optimizing the underlying protocols, approximations, or model structures~\cite{lou2021safenet,zhang2025cipherprune,zheng2023primer,zhang2026cryptogen,zhang2026securerouter}. Cheetah reduces communication in two-party neural-network inference~\cite{huang2022cheetah}; MPCFormer and related systems optimize MPC-based Transformer inference~\cite{li2022mpcformer}; HE-oriented systems such as THOR and Orion improve encrypted Transformer or neural-network execution through packing, rotation, and polynomial-evaluation optimizations~\cite{ebel2025orion,moon2024thor}; and PrivCirNet exploits structured model transformations to reduce private-inference cost~\cite{xu2024privcirnet}. These approaches improve specific backends or model designs, whereas our contribution is a hardware-aware compiler that uses measured primitive costs to choose a low-latency hybrid execution plan for each deployment. \rev{Closest to our setting, CostSphere~\cite{chen2025costsphere} assigns FHE/MPC protocols with an ILP over a linearized operator sequence using learned CPU cost models that adapt to network conditions; it approximates HE state by a consecutive-FHE count and does not model join compatibility or hardware changes. Matchmaker~\cite{jawalkar2025matchmaker} splits batches between LSS- and FSS-based MPC on CPU and GPU to maximize throughput, but stays within MPC and does not place individual operators. EncFormer~\cite{zhu2026encformer} improves the hybrid stack with stage-compatible FHE kernels and a complex CKKS--MPC conversion under hand-designed FHE/MPC boundaries. Their protocols and kernels can be profiled by MicroBench as additional actions; CipherFlow contributes deployment-specialized planning of placement, conversion, and ciphertext-state maintenance over branched graphs (Appendix~\ref{app:related}).}

\subsection{Threat Model}
We target standard cloud secure inference: the client holds private input $X$, the server holds model weights $W$, and only the agreed output is revealed. We assume a semi-honest two-party adversary, as widely adopted in secure inference~\cite{rathee2020cryptflow2,huang2022cheetah,zhang2025attestllm,zhang2024secure,kei2025shaft,knott2021crypten,xu2025breaking,lu2023bumblebee,pang2024bolt,xue2026robpi,santriaji2025dataseal}. Parties follow the protocol but may infer extra information from their local views; the client keeps key material locally, and the two parties do not collude with any external party. Malicious security, side channels, integrity, and verifiability are orthogonal to our compiler focus and can be incorporated through complementary techniques~\cite{moon2024thor,pang2024bolt}.

\subsection{Problem Formulation}
\label{sec:problem-formulation}
We formalize hardware-aware hybrid compilation as a constrained latency-minimization problem. The input consists of: \emph{(i)} a user model represented as a computation graph $G=(V,E)$, where $V=\{v_1,\ldots,v_N\}$ denotes the operators and $E$ their data dependencies; and \emph{(ii)} a deployment-specific cost profile $\mathcal{D}$ produced by MicroBench module under target hardware $\mathcal{H}$ and network conditions $(\mathrm{BW},\mathrm{RTT})$, which records the profiled compute, communication, and maintenance costs of HE/MPC operators, cross-domain conversions, and bootstrapping.

The output is a hybrid secure execution plan $P$ that assigns each operator $v_i \in V$ a cryptographic execution state
$
S_i = (d_i, l_i),
$
where $d_i \in \{\mathrm{HE}, \mathrm{MPC}\}$ is the execution domain and, when $d_i=\mathrm{HE}$, $l_i$ denotes the remaining HE level. In addition to operator states, the plan specifies the auxiliary actions required to realize valid transitions across the graph, including HE$\leftrightarrow$MPC conversions and HE maintenance actions such as bootstrapping.

A plan is valid if it preserves the functional semantics of $G$ and if its induced state assignments and transitions are feasible under the underlying cryptographic primitives. In particular, HE-domain operators must respect the level budget $l_i\geq 0$, and HE merge points must satisfy operand-compatibility constraints across incoming branches.

The objective is to find, among all valid plans, the one of minimum end-to-end inference latency:
\begin{equation}
P^*
\;=\;
\arg\min_{P \in \mathcal{P}(G,\mathcal{D})}
\;\;
\mathrm{Latency}_{\mathrm{e2e}}(G,P;\mathcal{D}),
\label{eq:problem}
\end{equation}
where $\mathcal{P}(G,\mathcal{D})$ denotes the set of valid plans for graph $G$ under deployment setting $\mathcal{D}$, and $\mathrm{Latency}_{\mathrm{e2e}}$ is computed by composing the per-state and per-transition costs profiled by \textit{MicroBench}.
The central technical problem is to solve Eq.~(1) efficiently on DNN graphs, where decisions are coupled by HE state evolution along paths and compatibility constraints across branches.

\section{CipherFlow Compiler Framework}
\begin{figure}[h]
    \centering
    \includegraphics[width=0.9\linewidth]{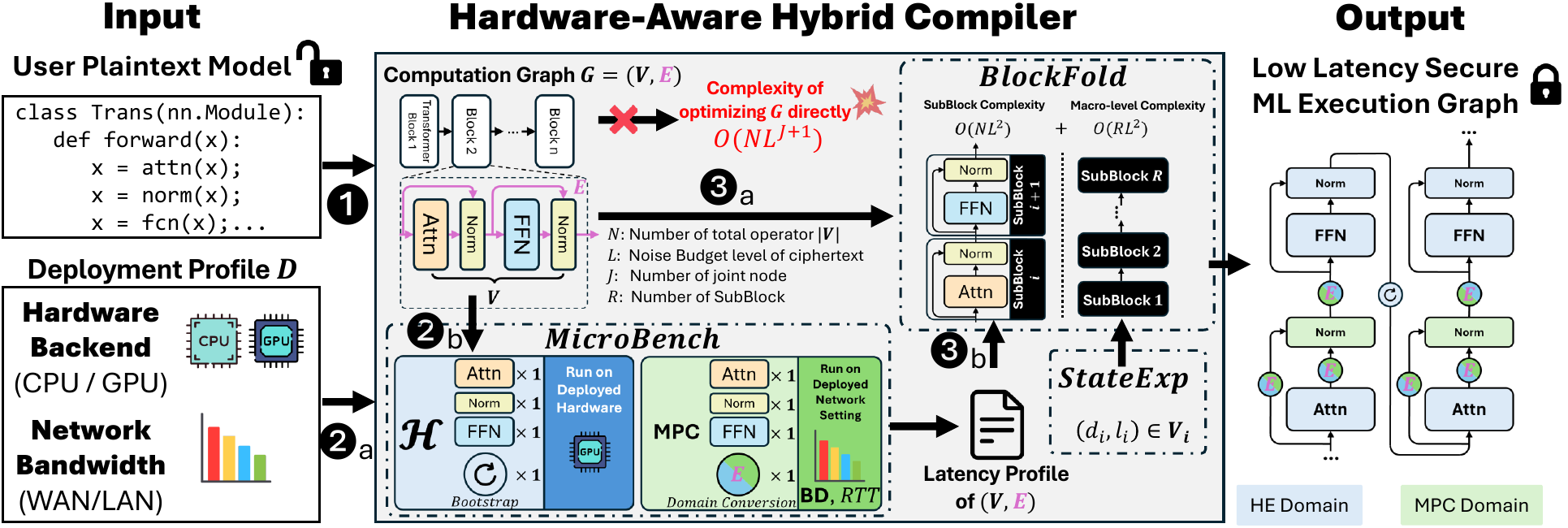}
    \vspace{-1.0 em}
    \caption{\textsc{CipherFlow} compilation pipeline. A model graph, supported cryptographic primitives, and deployment information determine a secure execution plan. \textit{MicroBench} supplies measured costs; state-expanded planning and \textit{BlockFold} select feasible operator, conversion, and maintenance actions. Circled numbers denote pipeline \emph{stages}: boxes sharing a number (e.g., \ding{193}a/\ding{193}b) are independent components whose outputs jointly feed the next stage.}
    \vspace{-1.0 em}
    \label{fig:overview}
\end{figure}

\textsc{CipherFlow} addresses the two challenges in Section~\ref{sec:intro} through a single compilation pipeline: it automates cryptographic execution decisions for \emph{programming usability} and jointly optimizes their deployment-specific costs for \emph{execution efficiency}. Figure~\ref{fig:overview} shows how a high-level AI computation becomes an executable hybrid HE--MPC program.

\textbf{Compiler inputs and output.}
The inputs are a model graph $G=(V,E)$, the available HE/MPC primitive implementations and conversion routines, and a target hardware and network configuration $(\mathcal{H},\mathrm{BW},\mathrm{RTT})$. The primitives determine the supported execution actions and their state transitions; \textit{MicroBench} supplies the deployment-specific cost profile $\mathcal{D}$ used in Eq.~\eqref{eq:problem}. The output is a secure execution plan $P$ that specifies backend operators, intermediate representations, HE$\leftrightarrow$MPC conversions, and ciphertext alignment and maintenance actions. These decisions make the plan executable using the supported primitives, beyond merely labeling operators as HE or MPC.

\textbf{Addressing programming usability.}
AI programmers specify tensor computations and dependencies while the compiler determines their cryptographic realization. State-expanded planning tracks representations and HE level budgets, and \textit{BlockFold} resolves compatibility across branches. The selected plan explicitly includes conversions, alignment, and refresh actions, so programmers do not manually schedule these operations. The compiler plans within the supported primitives' execution constraints and security assumptions; state feasibility is not a replacement for the underlying protocols' security guarantees.

\textbf{Addressing execution efficiency.}
\textit{MicroBench} measures the costs of available actions on the target deployment. State-expanded planning compares complete feasible sequences, jointly choosing operator placement, conversion points, and HE maintenance rather than selecting the cheapest operator in isolation. \textit{BlockFold} makes this optimization tractable on branched graphs by composing region summaries. Thus, the same model can be replanned for different hardware or networks without requiring the programmer to rewrite its cryptographic execution strategy.

\subsection{MicroBench: Deployment-Specific Cost Instantiation}
\label{sec:microbench}
\textsc{MicroBench} addresses execution efficiency by grounding the optimizer in deployment-specific measurements. It runs at compilation time, rather than per inference, and profiles HE/MPC execution, cross-domain conversions, and bootstrapping. This supplies the cost model automatically instead of requiring programmers to estimate which cryptographic primitive will be fastest.

The model is lowered to an operator-level DAG, from which \textsc{MicroBench} extracts distinct operator-shape patterns and profiles the available backends and conversion routines. Repeated tensor shapes and block structures in Transformer models allow measurements to be reused across operator instances. For each action, \textsc{MicroBench} records local compute time, communication volume, and round count when applicable. The planner composes these quantities with the target bandwidth and RTT to obtain the costs of execution, HE$\leftrightarrow$MPC conversion, and bootstrapping. The resulting profile $\mathcal{D}$ assigns costs to the legal actions used by the planner; the primitives' state-transition rules determine their feasibility. Additional profiling details are given in Appendix~\ref{app:microbench}.

\subsection{State-Expanded Planning}
\label{sec:state-expanded}
State-expanded planning connects programming usability to execution efficiency. It represents cryptographic states explicitly, so the compiler can insert the actions required for execution and compare their combined costs. An HE operator's feasibility depends on the inherited ciphertext state, especially its remaining level; an inexpensive placement can therefore cause expensive refreshes or conversions downstream. Planning over a \emph{state-expanded graph} captures these dependencies without asking the programmer to manage ciphertext levels manually.

\textbf{State-expanded nodes.}
A node is a triple $s=(i,d,l)$, pairing an operator index $i\in\{0,\ldots,N+1\}$ (with sentinel source and sink) with the state $S_i=(d,l)$ from Section~\ref{sec:problem-formulation}. Here $d\in\{\mathrm{HE},\mathrm{MPC}\}$ is the execution domain and $l$ is the remaining HE level. For MPC, the level component is inactive; we write $l=L$ by convention, consistent with an MPC$\rightarrow$HE conversion returning a fresh level budget. Including the inherited state in each node makes the residual planning cost locally additive along execution paths.

\textbf{Legal actions as state transitions.}
Edges represent five compiler actions: HE execution, MPC execution, HE$\rightarrow$MPC conversion, MPC$\rightarrow$HE conversion, and bootstrapping. Execution advances the operator index, conversion switches domains without consuming an operator, and bootstrapping refreshes the HE state. Only transitions satisfying the supported actions' state requirements are admissible. Edge weights come from \textsc{MicroBench}'s deployment-specific profile; the full transition rules are listed in Appendix~\ref{app:state_transitions}. Selecting a path therefore specifies both operator execution and the auxiliary protocol actions needed to make that execution feasible.

\textbf{Refresh by bootstrapping or by conversion.}
Both a direct bootstrap and an HE$\rightarrow$MPC$\rightarrow$HE round trip restore the ciphertext level. Bootstrapping is compute-only, whereas the round trip is communication-dominated, so their relative cost depends on the deployment. The planner chooses between these alternatives together with operator placement, relieving the programmer of manually selecting refresh points and mechanisms. Table~\ref{tab:refresh-crossover} illustrates how the preferred strategy changes with hardware and bandwidth.

\textbf{Sequential shortest-path reduction.}
On a sequential computation chain, a valid hybrid plan corresponds to a path from source state $s_{\mathrm{src}}$ to sink state $s_{\mathrm{dst}}$. Its modeled latency is the sum of the edge costs, so the minimum-cost path gives the optimal sequential plan under the profiled cost model. This jointly determines operator placement, conversions, and HE maintenance. The state-expanded graph contains $O(NL)$ nodes and edges on a chain, keeping planning tractable for the HE level ranges considered. Branched graphs additionally require compatibility across incoming paths; \textit{BlockFold} extends this formulation to those dependencies.

\subsection{BlockFold: Hierarchical Planning for Branched Graphs}
\label{sec:blockfold}
\textit{BlockFold} preserves the same programming abstraction for residual and attention branches: the input graph specifies their dependencies, and the compiler determines compatible cryptographic execution states at each join. Tracking every combination of branch states globally would make planning grow combinatorially. \textit{BlockFold} instead summarizes single-entry single-exit (SESE) regions and composes their boundary-state effects, supporting efficient planning while retaining the state constraints needed for execution.

\textbf{Step 1: SESE region extraction.}
The compiler identifies subgraphs with a unique entry point and a unique exit point, potentially containing internal branches and merges. Residual blocks, self-attention blocks, and multi-head attention structures provide such regions. Isolating them separates branch-local state interactions from the rest of the graph without requiring the programmer to annotate their HE/MPC placement.

\textbf{Step 2: Region transfer functions.}
For each SESE region $R$, \textit{BlockFold} solves a local planning problem over feasible execution, conversion, and maintenance actions. It enforces operand compatibility at HE merges (Section~\ref{sec:preliminaries}) through level-dropping or targeted bootstrapping. The region transfer function
\[
T_R(s^{R}_{\mathrm{src}}, s^{R}_{\mathrm{dst}})
=
\min_{P:\, s^{R}_{\mathrm{src}}\rightsquigarrow s^{R}_{\mathrm{dst}}}
\sum_{e\in P} c(e)
\]
records the minimum cost of traversing $R$ from input state $s^{R}_{\mathrm{src}}$ to output state $s^{R}_{\mathrm{dst}}$. Each entry summarizes feasible internal choices of HE/MPC placement, conversions, level alignment, and bootstrapping, exposing their boundary-state effect to the global planner.

\textbf{Step 3: Macro-level composition.}
Replacing each region with its transfer function yields an aggregate graph over region boundaries. A macro-level shortest-path search composes these summaries into a globally coordinated plan. This avoids tracking all branch-internal joint states across the full graph. The complexity analysis appears in Appendix~\ref{app:blockfold-complexity}, and the full algorithm is given in Appendices~\ref{appx:pseudocode}--\ref{app:state_transitions}.

\textbf{From an optimized plan to executable secure computation.}
After selecting the macro path, a backtracking pass expands the recorded region solutions onto the original graph. It materializes operator placement, cross-domain conversions, level alignment, and bootstrapping while retaining the graph's dependencies. The resulting plan $P$ specifies the backend operations and auxiliary actions required for execution. This completes the connection between the two challenges: the compiler takes over cryptographic execution decisions from the AI programmer, while measured costs and state-aware optimization determine an efficient realization for the target deployment.

\section{Evaluation}

We organize the evaluation around \emph{programming usability} and \emph{execution efficiency}. For programming usability, we examine HE-state feasibility and the scalability of automated plan generation. For execution efficiency, we examine deployment-specific inference latency and the contribution of cost modeling and joint planning. The usability evidence concerns compiler functionality and compilation cost; programmer effort is not measured. We first describe the experimental setup, then report execution latency, and finally group the component ablations by these two goals.

\subsection{Experimental Setup}
\label{subsec:set}
\textbf{Hardware.}
We evaluate on two target platforms: \textbf{CPU Platform:} A workstation equipped with an AMD Ryzen Threadripper PRO 3955WX 16-Core processor. \textbf{GPU Platform:} An HPC cluster equipped with an NVIDIA B200 GPU.

\textbf{Network and workload.}
We sweep bandwidth $\mathrm{BW} \in \{10~\mathrm{Mbps}, 100~\mathrm{Mbps}, 1~\mathrm{Gbps}, 3~\mathrm{Gbps}\}$ and $\mathrm{RTT} =\{1,20,40,80\}~\mathrm{ms}$, yielding 16 operating points per platform spanning constrained WAN to high-bandwidth LAN. We evaluate two workloads: BERT-base and ViT-B/16. BERT-base uses 12 Transformer blocks at sequence length 128, including attention QK/AV, out-projection,
FFN-Linear-1/2, Softmax, LayerNorm, GeLU, and residual additions; the compiler plans over 132 operator instances. ViT-B/16 uses 197 tokens (14$\times$14 patches plus a CLS token) with repeated self-attention, MLP blocks, LayerNorm, and residual connections, allowing us to evaluate whether the compiler generalizes beyond language Transformer inference.%

\textbf{Primitives.}
HE primitives are NEXUS CKKS on CPU and NEXUS-CUDA ($N=32768$) on GPU, both with a 20-level budget and bootstrapping. MPC primitives use SCI/SHAFT \cite{rathee2020cryptflow2,kei2025shaft,rathee2022secfloat} on CPU and SHAFT \cite{kei2025shaft} on GPU. HE and MPC operator latencies are measured as wall-clock time on the target hardware (Table~\ref{tab:microbench-entries}); SHAFT's bytes and rounds are its measured end-to-end totals apportioned to operators by compute time. HE$\leftrightarrow$MPC conversion is profiled by MicroBench under standard mask-and-decrypt and re-encryption protocols. 
We provide additional details on MicroBench, including its warm-up protocol, profiling procedureand representative operator-level cost entries, in Appendix~\ref{appx:measurement-methodology}.

\rev{\textbf{Baselines.} BumbleBee, BOLT, and BLB are applied as operator-to-domain placement rules (linear operators in HE, nonlinear operators in MPC) on our primitives; BOLT and BLB additionally receive a discount on conversion cost ($0.75\times$ and $0.5\times$) that models their batched and fused conversions. For BERT-base we also include the published BumbleBee end-to-end system measurement~\cite{lu2023bumblebee}, recomposed under each network setting, as a stronger system-level baseline. All models use static shapes (fixed sequence length); Section~\ref{sec:discussion} discusses dynamic shapes.}

\subsection{Execution Efficiency Across Deployments}
\begin{figure}[ht]
    \centering
    \includegraphics[width=0.9\linewidth]{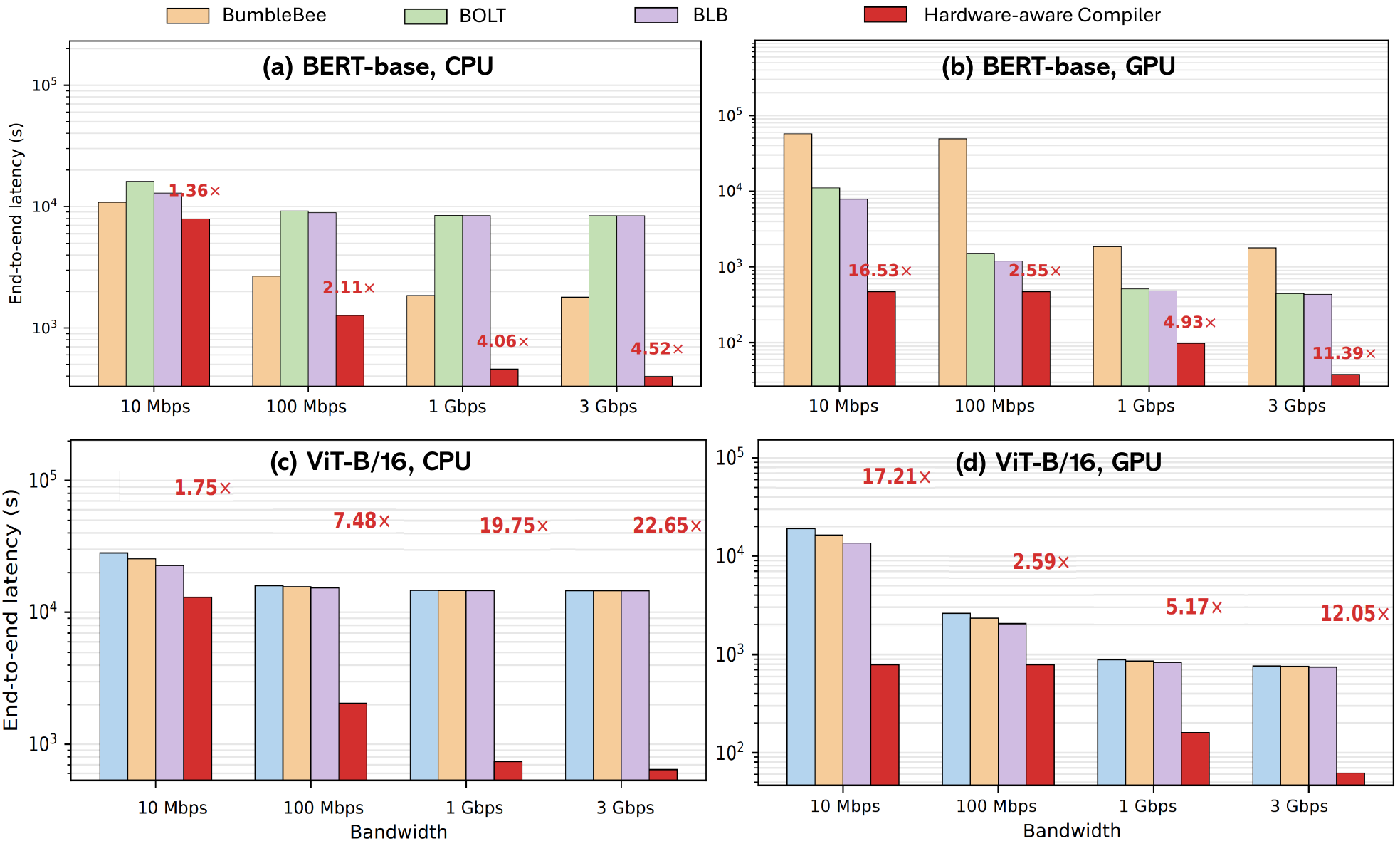}
    \vspace{-0.8 em}
    \caption{
    \rev{End-to-end latency of \textsc{CipherFlow} versus static hybrid baselines on BERT-base (a,b) and ViT-B/16 (c,d). For BERT-base, the orange bars are the published BumbleBee end-to-end system measurement recomposed under each network; BOLT and BLB (and all ViT baselines) apply the corresponding placement rules to our primitives. Panel (a) reports RTT $=1$\,ms, where the BumbleBee system is strongest; panels (b--d) report the RTT with the largest gap for each bandwidth. Speedup is relative to the fastest baseline in each group. The full $4\times4$ sweep, including placement-fair BumbleBee mappings, is in Appendix~\ref{appx:seqlen}.}
    }
    \label{fig:sota_placement}
\end{figure}

Figure~\ref{fig:sota_placement} (a,b) reports end-to-end BERT-base latency across four bandwidth settings on the CPU and GPU. Our compiler consistently matches or outperforms the best static placement baseline at every operating point. Apart from the published BumbleBee system measurement (orange bars in (a,b)), all bars within each group use the same profiled HE, MPC, conversion, and maintenance costs, so these improvements isolate the benefit of placement and conversion planning rather than differences in primitive implementations.%

The key observation is that the best static strategy changes with the deployment setting. On CPU, the relative ranking of static baselines varies as bandwidth increases, showing that a fixed HE/MPC assignment cannot robustly capture the trade-off between local cryptographic computation and communication. On GPU, the latency pattern differs from CPU, indicating that hardware acceleration shifts the HE-MPC cost balance and changes which placement is preferable. Therefore, the optimal BERT-base plan is both bandwidth-dependent and hardware-dependent. Appendix~\ref{app:placement-analysis} further visualizes the compiler-selected HE/MPC operator mix across bandwidth and RTT settings, showing how the same BERT-base graph induces different placements under CPU and GPU primitive profiles.

These results support the need for a compiler rather than a manually designed hybrid rule. A static strategy may perform well in one regime but become suboptimal when moved to another bandwidth or hardware setting. In contrast, our compiler uses the deployment-specific profile to select the lowest-latency plan for each operating point, explaining why the red bars remain at or below the best static baseline across Figure~\ref{fig:sota_placement}.

Figure~\ref{fig:sota_placement} (c,d) shows that the compiler provides even larger gains on ViT, where static hybrid mappings suffer from the longer token sequence. ViT-Base/16 uses $S=197$ tokens, compared with $S=128$ for BERT-base. This increases attention QK/AV costs quadratically with $S$ and FFN and conversion costs linearly with $S$. Static linHE+nlMPC-style mappings therefore pay amplified HE matmul latency while still inserting HE$\leftrightarrow$MPC conversions around nonlinear operators. In contrast, the compiler re-optimizes the entire placement under the measured deployment profile: when bandwidth is sufficiently high, it can move ViT to an all-MPC plan, eliminating conversion overhead and avoiding the enlarged HE attention bottleneck. 
BERT-base has a shorter sequence length, so HE matmuls remain more competitive and the compiler often selects a mixed plan, leading to smaller but still consistent gains. 

\subsection{Component Analysis: Usability and Efficiency}
The ablations connect the compiler components to execution efficiency through plan quality, and to programming usability through feasible and scalable automated planning. Table~\ref{tab:plan_ablation} reports the worst-case end-to-end latency ratio of each ablated planner relative to the full compiler across the evaluated BERT-base deployment settings. Table~\ref{tab:compile_ablation} reports the corresponding compilation cost on BERT-base.

\subsubsection{Execution Efficiency: Cost Modeling and Plan Quality}
\begin{table}[ht]
\centering
\caption{Plan-quality ablation and per-operator baselines.}
\label{tab:plan_ablation}
\begin{tabular}{lcc}
\toprule
Ablation & CPU latency ratio & GPU latency ratio \\
\midrule
Full & 1.00$\times$ & 1.00$\times$ \\
$-$MicroBench & 1.19$\times$ & 19.14$\times$ \\
$-$StateExp & 1.00$\times$ & 1.00$\times$ \\
$-$BlockFold & 1.00$\times$ & 1.00$\times$ \\
\rev{$-$Bootstrap} & \rev{1.00$\times$} & \rev{3.88$\times$} \\
\midrule
\rev{Greedy per-operator} & \rev{1.37$\times$} & \rev{5.75$\times$} \\
\rev{Random, best of 100} & \rev{10.92$\times$} & \rev{14.02$\times$} \\
\bottomrule
\end{tabular}
\vspace{-1em}
\end{table}

\textbf{MicroBench: deployment-specific cost modeling.}
Removing MicroBench replaces measured deployment-specific primitive costs with a fixed cost model. This directly degrades plan quality because the planner can no longer observe how the relative costs of HE execution, MPC execution, conversion, and maintenance change across CPU/GPU backends and network settings. As shown in Table~\ref{tab:plan_ablation}, this ablation increases the worst-case latency by 1.19$\times$ on CPU and by 19.14$\times$ on GPU. The much larger degradation on GPU confirms the central motivation of \textsc{CipherFlow}: hardware acceleration can substantially change the optimal hybrid strategy, so a static or deployment-agnostic cost model is insufficient.

\rev{\textbf{Per-operator baselines.} Greedy assigns each operator to the domain with lower local cost, followed by the necessary conversions and bootstraps. Random samples 100 uniform per-operator assignments. We evaluate all plans using the same cost model. Greedy performs no better than CipherFlow, particularly when adjacent operators favor different domains. On GPU at 10\,Mbps, Greedy assigns 24 of 132 operators to MPC, requiring 72 conversions and resulting in a $5.75\times$ slowdown compared with CipherFlow's all-HE plan. Even the best random assignments are $10.9$--$14.0\times$ slower (Table~\ref{tab:baselines-summary}). EncFormer's placement~\cite{zhu2026encformer}, which assigns linear operators to FHE and nonlinear operators to MPC with half-payload conversions, matches the BLB column under our profiles.}

\subsubsection{Programming Usability: Feasibility and Scalability}
\begin{table}[ht]
\centering
\caption{Compilation-time ablation on BERT-base.}
\label{tab:compile_ablation}
\begin{tabular}{lccc}
\toprule
Ablation & Compilation time & Ratio vs. Full & Method \\
\midrule
Full & 3.50 s & 1.00$\times$ & empirical \\
$-$MicroBench & 3.39 s & 0.96$\times$ & empirical \\
$-$StateExp & 2.36 s & 0.67$\times$ & empirical, infeasible \\
$-$BlockFold & $>$278 h & $>2.8\times10^5\times$ & search-space extrapolation \\
\bottomrule
\end{tabular}
\end{table}
\paragraph{State-expanded planning: HE-state feasibility.}
Removing state expansion makes the planner optimize only over operator domains, without explicitly tracking the remaining HE level along each execution path. This does not necessarily increase the repaired end-to-end latency in Table~\ref{tab:plan_ablation}, but the apparent result is misleading: without state expansion, the planner can select plans that violate HE feasibility constraints and therefore require after-the-fact repair. In contrast, the full compiler reasons over execution domain and HE state jointly, so placement, conversion, and maintenance decisions are optimized under feasibility constraints rather than patched after planning. The reduced compile time of the $-$StateExp variant in Table~\ref{tab:compile_ablation} therefore reflects a weaker and potentially infeasible optimization problem, not a better compiler.

\paragraph{BlockFold: scalable optimization over branched graphs.}

Removing BlockFold preserves the same local state-expanded formulation but applies it directly to the full branched graph. \rev{Its worst-case latency in Table~\ref{tab:plan_ablation} is unchanged, because BlockFold preserves the optimal boundary-state cost of each folded region; the chain relaxation used when BlockFold is removed can even report up to 11\% lower latency on GPU, but only by ignoring join-compatibility constraints.} Its benefit is scalability. As shown in Table~\ref{tab:compile_ablation}, the full compiler produces a BERT-base plan in 3.51 seconds, whereas the planner without BlockFold does not finish within the search budget and is extrapolated to require more than 278 hours, corresponding to a slowdown greater than $2.8\times10^5\times$. \rev{With BlockFold, compile time grows linearly with depth: 3.5, 7.3, 14.8, and 29.9\,s for 12, 24, 48, and 96 encoder blocks.} This result shows that BlockFold is necessary for extending HE-state-aware planning from sequential operator chains to Transformer-scale graphs with residual and attention joins.

The ablations connect the two goals to complementary compiler components. For execution efficiency, MicroBench provides the measured deployment-specific costs needed for low-latency plan quality. For programming usability, state-expanded planning enforces HE-state feasibility during optimization, and BlockFold makes automated state-aware planning scalable on branched neural-network graphs.

\section{Discussion}
\label{sec:discussion}
\noindent\textbf{Programming usability through compilation.}
CipherFlow separates model development from cryptographic execution planning. AI programmers specify a model graph; the compiler combines available HE/MPC primitives with the target hardware and network configuration to construct an operator-level secure execution plan. State-expanded planning makes conversions, HE maintenance, and state compatibility explicit, while BlockFold makes planning scalable across branched graphs. The evaluation examines this compiler-level usability through feasibility and compilation cost; programmer effort is not measured. Security relies on the supported primitives and their assumptions.

\noindent\textbf{Execution efficiency through deployment-specific planning.}
Our results suggest that HE--MPC placement is better handled as a deployment-specific compilation problem than through fixed operator-mapping rules. Using measured primitive costs from MicroBench, CipherFlow jointly selects operator domains, conversions, and HE maintenance operations to adapt to different hardware and network conditions. Thus, the same abstraction that automates cryptographic decisions also allows their combined execution cost to be optimized for the target deployment.

\noindent\textbf{Broader objectives and batching.}
Although we optimize latency, other additive objectives, such as monetary cost or weighted latency and communication volume, can be incorporated by changing edge weights. Hard communication budgets require tracking cumulative volume in the planning state. Batching may also change the preferred placement: CKKS SIMD packing can amortize HE computation across requests, while MPC communication scales with the processed data. Batch-aware planning can use profiled batch costs, but jointly optimizing packing and placement, as explored for packing in Coyote~\cite{malik2023coyote}, remains open.

\noindent\textbf{An extensible platform for private inference.}
CipherFlow is open source
(\href{https://github.com/loopleap-lab/CipherFlow}{github.com/loopleap-lab/CipherFlow}).
New HE/MPC libraries and kernels can be integrated through registered methods and measured costs, allowing the planner to benefit from advances in either backend while retaining the model-level programming interface.
Future work includes generative models and encrypted execution.

\section{Acknowledgment}
This work was supported in part by the National Science Foundation (NSF) under Award Nos. 2523407 and 2413232. Any opinions, findings, conclusions, or recommendations expressed in this material are those of the authors and do not necessarily reflect the views of NSF.

\clearpage
\bibliographystyle{unsrtnat}
{\small
\bibliography{main}
}

\clearpage
\appendix
\section{Related works.}\label{app:related} Hybrid HE-MPC systems~\cite{juvekar2018gazelle,riazi2018chameleon,xu2025breaking,lu2023bumblebee,pang2024bolt} exploit the complementary costs of HE and MPC for private inference, while secure-computation compilers and cost-modeling frameworks such as EzPC~\cite{chandran2019ezpc}, MP2ML~\cite{boemer2020mp2ml} and nGraph-HE2~\cite{boemer2019ngraph} show the importance of programmability and deployment-aware costs. Our work is complementary: rather than optimizing a single backend or using a fixed hybrid placement policy, we compile HE/MPC placement, bidirectional conversion, and HE maintenance decisions under a measured deployment profile.

\rev{\textbf{CostSphere}~\cite{chen2025costsphere} is the closest prior system. It formulates protocol assignment as an ILP over a topologically linearized sequence of TOSA-level operators: conversion costs are charged only between adjacent positions, so cross-branch compatibility at residual and attention joins is not modeled. It approximates HE state by counting consecutive FHE operators and triggering a bootstrap after a threshold, whereas level consumption is heterogeneous (16 levels for Softmax vs.\ 1 for a matrix multiplication). Its costs come from offline regression models trained on one CPU platform with bandwidth features in the MPC model only, so the hardware dimension that reverses our optimal placement (Figure~\ref{fig:mix_heatmap}) is absent. CipherFlow instead plans exact per-level state on the branched graph with deployment-time measurements on CPU or GPU, in polynomial time.

\textbf{Matchmaker}~\cite{jawalkar2025matchmaker} shares the view that no single configuration is optimal across deployments. It operates within 2PC MPC in the preprocessing model, choosing the fraction of a batch executed by a CPU-optimized LSS suite versus a GPU-based FSS suite to minimize batch completion time; it places no individual operators and has no HE domain. Its protocol suite could serve as an MPC backend profiled by MicroBench.

\textbf{EncFormer}~\cite{zhu2026encformer}, contemporaneous with this work, improves the hybrid stack itself (stage-compatible packing, a complex CKKS--MPC conversion, and communication-efficient nonlinear protocols) and selects FHE/MPC boundaries with offline cost equations over a small hand-enumerated design space. Its authors list cost-model-guided compilers that reduce hand-crafted placement as future work, which is the problem CipherFlow addresses. Its placement is the linear-in-HE, nonlinear-in-MPC family evaluated in Table~\ref{tab:baselines-summary}.

\textbf{Coyote}~\cite{malik2023coyote} vectorizes encrypted arithmetic circuits within HE; batching-aware packing of this kind is complementary to cross-domain placement (Section~\ref{sec:discussion}).}

\section{MicroBench Profiling Details}
\label{app:microbench}
\textsc{MicroBench} is executed once for a target deployment profile before planning. It is not invoked per inference request. Given a model graph and a target backend, \textsc{MicroBench} extracts the distinct operator-shape patterns that appear in the graph and measures the secure primitive latency for each supported execution action. The resulting profile is cached and reused by the planner for all inputs under the same backend configuration. 

\textsc{MicroBench} profiles five action types: (i) HE execution, (ii) MPC execution, (iii) HE$\rightarrow$MPC conversion, (iv) MPC$\rightarrow$HE conversion, and (v) HE maintenance through bootstrapping. (Detail in Table \ref{tab:state_transitions}. The two conversion directions are measured separately because their compute, communication, and packing patterns are generally asymmetric.

\begin{table}[ht]
\centering\small
\caption{Refresh cost for a $[1,128,768]$ tensor (RTT $=20$\,ms): a direct bootstrap is compute-only, whereas the conversion round trip is communication-dominated, so the cheaper mechanism changes with the deployment.}
\label{tab:refresh-crossover}
\begin{tabular}{lcccc}
\toprule
Refresh mechanism & 10\,Mbps & 100\,Mbps & 1\,Gbps & 3\,Gbps \\
\midrule
HE$\rightarrow$MPC$\rightarrow$HE round trip & 63\,s & 6.33\,s & 0.672\,s & 0.252\,s \\ \midrule
Direct bootstrap, CPU (51.6\,s) & \textbf{bootstrap} & round trip & round trip & round trip \\
Direct bootstrap, GPU (1.69\,s) & \textbf{bootstrap} & \textbf{bootstrap} & round trip & round trip \\
\bottomrule
\end{tabular}
\end{table}

\begin{table}[ht]
\centering\small
\caption{Removing the direct-bootstrap transition (BERT-base). The planner may still refresh ciphertexts through HE$\rightarrow$MPC$\rightarrow$HE round trips.}
\label{tab:nobootstrap}
\begin{tabular}{lccccc}
\toprule
Platform & BW & RTT & Full & $-$Bootstrap & Slowdown \\
\midrule
GPU & 10\,Mbps & 1\,ms & 472\,s & 1.83\,ks & $3.87\times$ \\
GPU & 10\,Mbps & 20\,ms & 472\,s & 1.83\,ks & $3.87\times$ \\
GPU & 10\,Mbps & 40\,ms & 472\,s & 1.83\,ks & $3.87\times$ \\
GPU & 10\,Mbps & 80\,ms & 472\,s & 1.83\,ks & $3.88\times$ \\
GPU & 100\,Mbps & 1\,ms & 472\,s & 582\,s & $1.23\times$ \\
GPU & 100\,Mbps & 20\,ms & 472\,s & 583\,s & $1.23\times$ \\
GPU & 100\,Mbps & 40\,ms & 472\,s & 584\,s & $1.24\times$ \\
GPU & 100\,Mbps & 80\,ms & 472\,s & 586\,s & $1.24\times$ \\
\midrule
CPU (all 16 points) & \multicolumn{2}{c}{all} & \multicolumn{2}{c}{unchanged} & $1.00\times$ \\
GPU, $\geq$1\,Gbps & \multicolumn{2}{c}{8 points} & \multicolumn{2}{c}{unchanged} & $1.00\times$ \\
\bottomrule
\end{tabular}
\end{table}

\rev{\textbf{Bootstrapping versus conversion-based refresh.} A direct bootstrap involves no communication or interaction and is bound by compute hardware, whereas an HE$\rightarrow$MPC$\rightarrow$HE round trip is bound by bandwidth and RTT. Table~\ref{tab:refresh-crossover} shows that the cheaper mechanism flips within the evaluated grid on both platforms, and Table~\ref{tab:nobootstrap} shows the cost of removing direct bootstrapping from the planner.}

\textbf{Sensitivity to HE Level-Dependent Costs.} Our HE profiles use NEXUS kernel costs measured at fixed levels. Table~\ref{tab:level-profile} measures these kernels across a 20-level chain on CPU and GPU (Section~\ref{subsec:set}); for $N=2^{13}$ and $2^{15}$, these chains exceed the HE standard's 128-bit security bound and are used only for cost profiling. Each additional level increases non-linear kernel latency by 8--14\%, while MatMul latency grows by $5.4\times$ on GPU and $20\times$ on CPU from level 1 to 20. Re-planning all $4\times4$ configurations with per-level costs and free modulus switching preserves HE/MPC placement across all 64 settings. CPU plans and predicted latencies remain unchanged, while GPU plans increase the median bootstrap count from 8.5 to 24 to execute linear layers at lower levels. Predicted GPU latency increases by at most 13\% for BERT-base and 8\% for ViT-B/16, with speedups over the best hybrid baseline of 2.16--14.69$\times$ and 2.34--15.94$\times$, respectively (versus 2.42--16.53$\times$ and 2.52--17.21$\times$ under single-level costs). Without re-planning, the original plans are up to $2.5\times$ and $2.8\times$ slower under per-level costs. Accounting for the higher level consumption of NEXUS non-linear kernels (16/18/18 versus the assumed 8/3/4 for Softmax/LayerNorm/GeLU) leaves CPU plans unchanged and yields GPU speedups of 2.15--14.59$\times$ and 2.33--15.82$\times$ for BERT-base and ViT-B/16, respectively.

\begin{table}[ht]
\centering\small
\setlength{\tabcolsep}{4pt}
\caption{Level-dependent HE cost of the NEXUS kernels behind the HE profiles. Each kernel keeps ring dimension and prime sizes on a 20-level chain; its input is encrypted at the top level and mod-switched to level $\ell$ before the timed region. $\delta$: levels the planner charges / levels the kernel consumes. Columns give the runtime at input level $\ell$ relative to the level $\ell_{\mathrm{ref}}$ at which the profile cost was measured (--: below the kernel's depth).}
\label{tab:level-profile}
\begin{tabular}{l c c c cccccccc}
\toprule
Kernel & Device & $\delta$ & $\ell_{\mathrm{ref}}$ & $1$ & $2$ & $4$ & $8$ & $12$ & $16$ & $18$ & $20$ \\
\midrule
MatMul & CPU & 1 / 1 & 1 & 1.00 & 1.55 & 2.84 & 6.06 & 10.01 & 14.87 &  & 20.42 \\
Softmax & CPU & 8 / 16 & 18 & -- & -- & -- & -- & -- & 0.74 & 1.00 & 1.29 \\
LayerNorm & CPU & 3 / 18 & 18 & -- & -- & -- & -- & -- & -- & 1.00 & 1.23 \\
GeLU & CPU & 4 / 18 & 18 & -- & -- & -- & -- & -- & -- & 1.00 & 1.27 \\
\midrule
MatMul & GPU & 1 / 1 & 1 & 1.00 & 1.17 & 1.67 & 2.70 & 3.70 & 4.49 &  & 5.36 \\
Softmax & GPU & 8 / 16 & 16 & -- & -- & -- & -- & -- & 1.00 & 1.20 & 1.39 \\
LayerNorm & GPU & 3 / 18 & 18 & -- & -- & -- & -- & -- & -- & 1.00 & 1.17 \\
GeLU & GPU & 4 / 18 & 18 & -- & -- & -- & -- & -- & -- & 1.00 & 1.15 \\
\bottomrule
\end{tabular}
\end{table}

\section{Limitation}
CipherFlow currently optimizes single-request execution under fixed deployment profiles and does not account for concurrency, batching, queueing, or resource contention. Throughput optimization therefore requires a serving-level cost model. HE costs are profiled at a single level per operator; level-dependent costs can be incorporated into the planner but have not been measured. The current implementation also assumes static computation graphs and tensor shapes. Unseen shapes can be profiled and cached, but autoregressive decoding with evolving KV caches and data-dependent routing in mixture-of-experts models require additional support. Finally, our security model assumes semi-honest two-party execution.

\section{Compiler Pseudocode}
\label{appx:pseudocode}

This appendix gives the algorithmic skeleton of the hardware-aware hybrid HE-MPC compiler. Algorithm~\ref{alg:compile} is the top-level driver: it consumes a model graph $G$ and a deployment setting $\mathcal{D}$, calls \textsc{MicroBench} to instantiate a deployment-specific cost profile $\Pi$, decomposes $G$ into SubBlock regions, computes a region transfer function $T_R$ for each region by exact state-expanded planning, and finally composes regions through a macro-level shortest-path search over region boundary states. Algorithm~\ref{alg:state-sssp} details the inner state-expanded shortest-path subroutine used in two places: \emph{(i)} computing $T_R$ for a SubBlock region, and \emph{(ii)} composing across regions in the macro graph.

\begin{algorithm}[ht]
\caption{\textsc{Compile}: Hardware-Aware Hybrid HE--MPC Compilation}
\label{alg:compile}
\begin{algorithmic}[1]
\Require Model graph $G=(V,E)$; deployment $\mathcal{D}=
(\mathcal{H},\mathrm{BW},\mathrm{RTT})$; HE level budget $L$
\Ensure Optimized hybrid execution plan $P^*$
\Statex
\State $\Pi \gets \textsc{MicroBench}(G,\mathcal{H})$
\State $w \gets \textsc{ComposeCost}(\Pi,\mathrm{BW},\mathrm{RTT})$
\State $\mathcal{R} \gets \textsc{ExtractSubBlock}(G)$
\Statex
\ForAll{$R \in \mathcal{R}$}
    \State $G_R^{\mathrm{state}} \gets
        \textsc{BuildStateGraph}(R,L,w)$
    \ForAll{$(s_{\mathrm{in}},s_{\mathrm{out}})
        \in \mathcal{B}_R$}
        \State $(T_R,\pi_R)[s_{\mathrm{in}},s_{\mathrm{out}}]
            \gets \textsc{StateEXP}(
            G_R^{\mathrm{state}},s_{\mathrm{in}},s_{\mathrm{out}})$
    \EndFor
\EndFor
\Statex
\State $G^{\mathrm{macro}} \gets
    \textsc{BuildMacroGraph}(G,\mathcal{R},\{T_R\})$
\State $(C^*,\pi^*) \gets
    \textsc{StateEXP}(G^{\mathrm{macro}},
    s_{\mathrm{src}},s_{\mathrm{dst}})$
\State $P^* \gets
    \textsc{Materialize}(\pi^*,\{\pi_R\})$
\State \Return $P^*$
\end{algorithmic}
\end{algorithm}

\begin{algorithm}[ht]
\caption{\textsc{StateEXP}: State-Expanded Shortest Path}
\label{alg:state-sssp}
\begin{algorithmic}[1]
\Require State graph $\mathcal{G}=(\mathcal{S},\mathcal{T},w)$;
source $s_{\mathrm{src}}$; target $s_{\mathrm{dst}}$
\Ensure Minimum cost $C^*$ and predecessor map $\pi$
\State $C[s] \gets \infty$ for all $s\in\mathcal{S}$
\State $C[s_{\mathrm{src}}] \gets 0$;\quad $\pi \gets \emptyset$
\State $Q \gets \{(0,s_{\mathrm{src}})\}$
\While{$Q \neq \emptyset$}
    \State $(c,s) \gets \textsc{ExtractMin}(Q)$
    \State \textbf{if} $c>C[s]$ \textbf{then continue}
    \State \textbf{if} $s=s_{\mathrm{dst}}$ \textbf{then break}
    \ForAll{$(s,s')\in\mathrm{Out}(s)$}
        \State \textbf{if} $\neg\textsc{Feasible}(s,s')$
            \textbf{then continue}
        \State $c' \gets c+w(s,s')$
        \If{$c'<C[s']$}
            \State $C[s']\gets c'$;\quad $\pi[s']\gets s$
            \State $\textsc{Push}(Q,(c',s'))$
        \EndIf
    \EndFor
\EndWhile
\State \Return $C[s_{\mathrm{dst}}],\pi$
\end{algorithmic}
\end{algorithm}

\section{StateExp}\label{app:state_transitions}
\rev{\textbf{State.} A state is $s=(i,d,\ell)$: operator index $i$ along the linearized spine, domain $d\in\{\mathrm{HE},\mathrm{MPC}\}$, and remaining level $\ell\in\{0,\dots,L\}$ (canonically $\ell=L$ for MPC). Table~\ref{tab:state_transitions} lists the five transitions. HE execution of $v_i$ is feasible only if $\ell\ge\delta_i$ and $v_i$'s HE kernel contract admits its input shape; otherwise the transition is absent, so infeasible placements are never generated. Conversions and bootstrapping do not advance $i$; MPC execution never changes $\ell$.

\textbf{Join handling.} A SESE region consists of a fork $u$, a main branch, a skip branch, and a join $v$ (a residual addition). For each entry state, BlockFold runs one state-expanded shortest-path search per branch, obtaining the cost of every reachable end state of each branch. For every pair of branch end states and each merge domain $d_v\in\{\mathrm{HE},\mathrm{MPC}\}$ it (i) converts any branch whose domain differs from $d_v$; (ii) if $d_v=\mathrm{HE}$, aligns levels by one of four strategies: drop the higher-level operand to the lower level (free in CKKS), bootstrap the main branch, bootstrap the skip branch, or bootstrap both; (iii) executes $v$ in $d_v$; and (iv) converts to the requested exit state. The minimum over all combinations is $T_R(s_{\mathrm{src}},s_{\mathrm{dst}})$, and the argmin is stored for backtracking. Algorithm~\ref{alg:blockfold} summarizes the procedure.}

\textbf{Worked example (BERT encoder block).} The block has 11 operators and two SESE regions: $R_1$ (fork QK, main branch Softmax$\rightarrow$AV$\rightarrow$Out-Proj, skip edge QK$\rightarrow$Add$_1$) and $R_2$ (fork LN$_1$, main branch FFN$_1\rightarrow$GeLU$\rightarrow$FFN$_2$, skip edge LN$_1\rightarrow$Add$_2$). With $L=20$ there are $|\mathcal{B}|=22$ boundary states (21 HE levels and MPC), so each region stores a $22\times22$ table. The macro spine is QK, $R_1$, LN$_1$, $R_2$, LN$_2$, and 12 blocks are chained by LN$_2\rightarrow$QK. On GPU at 10\,Mbps every table entry that stays in HE is cheapest (bootstrapping 1.7\,s vs.\ a 63\,s round trip), and the planner outputs an all-HE plan with 17 bootstraps; on CPU at 3\,Gbps the MPC-entry/MPC-exit entries dominate and the plan is all-MPC. Table~\ref{tab:worked-example} traces one block at an intermediate setting (GPU, 50\,Mbps/20\,ms, input arriving secret-shared) where a single block uses every action. QK stays in MPC, where its input already is; its output is converted to HE because MPC Softmax would transfer 106\,MiB. At Add$_1$ the skip operand, still in MPC, is converted and dropped for free to the main branch's level 10. After LN$_1$ the rest of the block needs $1{+}4{+}1{+}3=9$ levels but only 7 remain, so the planner bootstraps (1.7\,s) rather than taking an HE$\rightarrow$MPC$\rightarrow$HE round trip (12.6\,s). At Add$_2$ both operands are aligned to the skip branch's level 7, and LN$_2$ leaves the block at level 4.

\begin{table}[ht]
\centering\small
\setlength{\tabcolsep}{4pt}
\caption{Plan of one BERT-base encoder block (GPU profile, 50\,Mbps / 20\,ms, input arriving in MPC), one row per planner action with the state $(d,\ell)$ before and after it (levels without optional free drops). $R_1$, $R_2$: the two SESE regions; QK, LN$_1$, LN$_2$ lie on the macro spine.}
\label{tab:worked-example}
\begin{tabular}{c l l c c r}
\toprule
Region & Operator & Action & Before & After & Cost (s) \\
\midrule
 & QK & MPC exec & (MPC) & (MPC) & 3.28 \\
$R_1$ & input of Softmax & MPC$\to$HE & (MPC) & (HE, 20) & 12.61 \\
$R_1$ & Softmax & HE exec & (HE, 20) & (HE, 12) & 0.26 \\
$R_1$ & AV & HE exec & (HE, 12) & (HE, 11) & 3.75 \\
$R_1$ & Out-Proj & HE exec & (HE, 11) & (HE, 10) & 3.75 \\
$R_1$ & skip operand of Add$_1$ & MPC$\to$HE & (MPC) & (HE, 20) & 6.31 \\
$R_1$ & Add$_1$ & HE exec & (HE, 10) & (HE, 10) & 0.00 \\
 & LN$_1$ & HE exec & (HE, 10) & (HE, 7) & 0.72 \\
$R_2$ & input of FFN$_1$ & bootstrap & (HE, 7) & (HE, 20) & 1.69 \\
$R_2$ & FFN$_1$ & HE exec & (HE, 20) & (HE, 19) & 15.00 \\
$R_2$ & GeLU & HE exec & (HE, 19) & (HE, 15) & 1.51 \\
$R_2$ & FFN$_2$ & HE exec & (HE, 15) & (HE, 14) & 3.75 \\
$R_2$ & Add$_2$ & HE exec & (HE, 7) & (HE, 7) & 0.00 \\
 & LN$_2$ & HE exec & (HE, 7) & (HE, 4) & 0.72 \\
\midrule
\multicolumn{5}{l}{Total (planner)} & 53.4 \\
\bottomrule
\end{tabular}
\end{table}

\begin{algorithm}[ht]
\caption{\textsc{BlockFold}: Region Transfer and Macro Composition}
\label{alg:blockfold}
\begin{algorithmic}[1]
\Require Graph $G$; edge weights $w$; HE level budget $L$
\Ensure Region transfer tables $\{T_R\}$ and macro graph $G^{\mathrm{macro}}$
\State $\mathcal{B} \gets
    \{(\mathrm{HE},\ell)\}_{\ell=0}^{L}
    \cup \{(\mathrm{MPC},L)\}$
\State $\mathcal{R} \gets \textsc{ExtractSESE}(G)$
\Statex
\ForAll{$R=(u,m,k,v)\in\mathcal{R}$}
    \ForAll{$s\in\mathcal{B}$}
        \State $D_m \gets \textsc{StateEXP}(m,s)$
        \State $D_k \gets \textsc{StateEXP}(k,s)$
        \ForAll{$s'\in\mathcal{B}$}
            \State $(T_R[s,s'],\pi_R[s,s'])
                \gets \textsc{JoinOpt}(D_m,D_k,v,s')$
        \EndFor
    \EndFor
\EndFor
\Statex
\State $G^{\mathrm{macro}} \gets
    \textsc{Collapse}(G,\mathcal{R},\{T_R\})$
\State \Return $\{T_R,\pi_R\},G^{\mathrm{macro}}$
\end{algorithmic}
\end{algorithm}

\begin{table}[ht]
    \centering
    \small
    \caption{State transitions in the expanded graph at operator $v_i$.}
    \label{tab:state_transitions}
    \begin{tabular}{p{0.22\linewidth} p{0.40\linewidth} p{0.14\linewidth} p{0.14\linewidth}}
        \toprule
        Transition type & State transition & Constraint & Cost \\
        \midrule
        HE execution & $(i,\mathrm{HE},l) \rightarrow (i{+}1,\mathrm{HE},l-\delta_i)$ & $l \ge \delta_i$ & $c_i^{\mathrm{HE}}$ \\
        MPC execution & $(i,\mathrm{MPC},L) \rightarrow (i{+}1,\mathrm{MPC},L)$ & - & $c_i^{\mathrm{MPC}}$ \\
        HE$\rightarrow$MPC conversion & $(i,\mathrm{HE},l) \rightarrow (i,\mathrm{MPC},L)$ & - & $c_i^{\mathrm{HE}\rightarrow\mathrm{MPC}}$ \\
        MPC$\rightarrow$HE conversion & $(i,\mathrm{MPC},L) \rightarrow (i,\mathrm{HE},L)$ & - & $c_i^{\mathrm{MPC}\rightarrow\mathrm{HE}}$ \\
        Bootstrapping & $(i,\mathrm{HE},l) \rightarrow (i,\mathrm{HE},L)$ & - & $c_i^{\mathrm{BS}}$ \\
        \bottomrule
    \end{tabular}
\end{table}

\section{Complexity Analysis of BlockFold}
\label{app:blockfold-complexity}
We analyze the planning complexity before and after BlockFold. Let
\(G=(V,E)\) be the computation graph with \(N=|V|\) operators. Let \(L\) denote
the number of HE level states tracked by the planner. Since each operator can be
assigned to either HE or MPC, and HE states additionally carry a level index, we
write \(q=O(L)\) for the number of cryptographic execution states per operator.
Let \(J\) be the number of join nodes in the graph, and let \(k_j\) denote the
number of incoming branches of the \(j\)-th join.

\paragraph{Planning without BlockFold.}
Without BlockFold, the planner performs global state tracking over the original
branched graph. At a join node, the planner must enforce compatibility among all
incoming branch states: the domains must agree, and if the join is executed in
HE, the incoming ciphertext levels must be aligned or made compatible through
maintenance actions. If the \(j\)-th join has \(k_j\) incoming branches and each
branch can end in one of \(q\) states, then the join induces \(O(q^{k_j})\)
joint state combinations.

A sequential chain with \(N\) operators would require \(O(Nq)\) state nodes.
However, each join introduces additional cross-branch state combinations. Since
one branch state is already represented in the sequential state propagation,
the \(j\)-th join contributes an additional multiplicative factor of
\(q^{k_j-1}\). Therefore, the worst-case global planning search space without
BlockFold is

\[
O\!\left(
Nq \prod_{j=1}^{J} q^{k_j-1}
\right).
\]

When all joins have the same fan-in \(k\), this simplifies to

\[
O\!\left(
N q^{1+J(k-1)}
\right).
\]

Using \(q=O(L)\), the complexity becomes

\[
O\!\left(
N L^{1+J(k-1)}
\right).
\]

Thus, even for binary joins (\(k=2\)), naive global planning can grow as
\(O(NL^{1+J})\), which becomes impractical for residual and attention-heavy
models with many joins.

\paragraph{Planning with BlockFold.}
BlockFold decomposes the graph into \(R\) single-entry single-exit (SubBlock)
regions. Let \(n_r\) be the number of operators in the \(r\)-th region, so that
\(\sum_{r=1}^{R} n_r = N\). Instead of exposing all internal branch states to
the global planner, BlockFold summarizes each region by a boundary-state
transfer function. For each pair of entry and exit states, the planner computes
the minimum internal cost of the region.

Each region has \(q\) possible entry states and \(q\) possible exit states,
resulting in \(O(q^2)\) boundary-state pairs. For a region of size \(n_r\), the
state-expanded shortest-path computation over all boundary pairs costs
\(O(n_r q^2)\). Therefore, the total cost of local region planning is

\[
O\!\left(
\sum_{r=1}^{R} n_r q^2
\right).
\]

After each region is summarized by its transfer function, the macro-level graph
contains \(R\) region nodes, each with \(O(q^2)\) transfer edges between
boundary states. Macro-level composition therefore costs

\[
O(Rq^2).
\]

Combining local region planning and macro-level composition, the overall
planning complexity with BlockFold is

\[
O\!\left(
\sum_{r=1}^{R} n_r q^2 + Rq^2
\right).
\]

Since \(\sum_{r=1}^{R} n_r=N\), this becomes

\[
O\!\left(
(N+R)q^2
\right).
\]

Substituting \(q=O(L)\), we obtain

\[
O\!\left(
(N+R)L^2
\right).
\]

Because \(R \leq N\) for a valid decomposition into non-empty regions, the
planning complexity is bounded by

\[
O(NL^2).
\]

Therefore, BlockFold reduces planning from exponential dependence on the number
of joins in the worst case,
\(O(NL^{1+J(k-1)})\), to polynomial dependence on the number of HE states,
\(O(NL^2)\), while preserving exact optimization within each SubBlock region.

\section{Operator Placement under Hardware and Network Variation}
\label{app:placement-analysis}
\begin{figure}[ht]
    \centering
    \includegraphics[width=\linewidth]{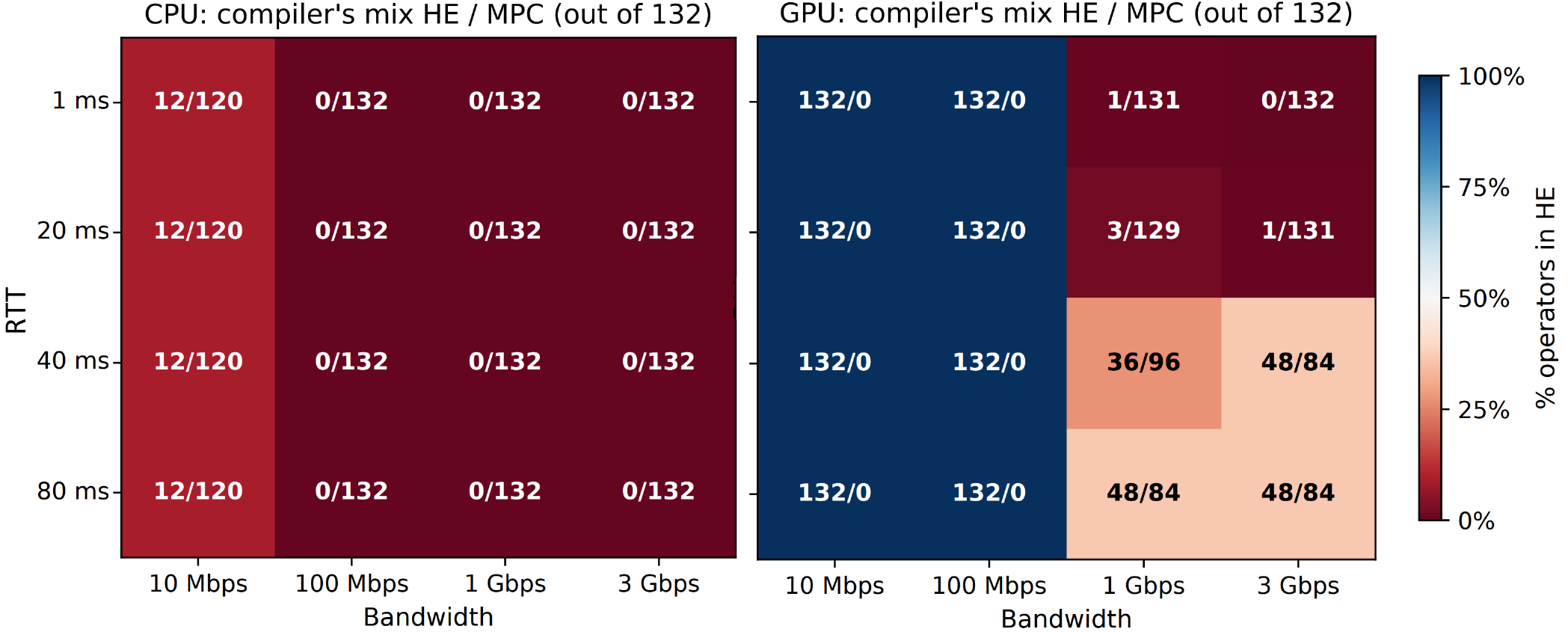}
    \caption{\textbf{Compiler-selected HE/MPC operator mix for BERT-base.} Each cell reports the number of operators placed in HE/MPC out of 132 operators, and color indicates the percentage of operators in HE. CPU and GPU exhibit opposite placement trends: CPU favors MPC once bandwidth is moderate, while GPU favors HE under bandwidth-constrained settings and introduces MPC only when communication becomes sufficiently cheap.}
    \label{fig:mix_heatmap}
\end{figure}
Figure~\ref{fig:mix_heatmap} reports the compiler-selected HE/MPC operator mix for BERT-base across bandwidth and RTT settings. The two platforms exhibit opposite placement patterns. On CPU, the compiler selects an MPC-dominant plan: at 10 Mbps it keeps only 12 of 132 operators in HE, and from 100 Mbps to 3 Gbps it selects a pure-MPC plan across all RTT values. This indicates that, under the CPU primitive profile, HE computation is sufficiently expensive that even communication-sensitive MPC becomes preferable once bandwidth is moderate.

GPU shows a qualitatively different pattern. At 10 and 100 Mbps, the compiler selects pure HE across all RTT values, reflecting the strong acceleration of HE operators and the high cost of MPC communication under constrained bandwidth. As bandwidth increases to 1-3 Gbps, the compiler begins to introduce MPC and mixed plans, with the exact mix depending jointly on bandwidth, RTT, and operator-level costs.

\section{Full Deployment Sweep}\label{appx:seqlen}
This appendix reproduces the complete $4\!\times\!4$ (bandwidth $\times$ RTT) sweep for both Transformer benchmarks side by side. BERT-base uses 12 encoder blocks at $S{=}128$ (132 operator instances); ViT-Base/16 uses 12 encoder blocks at $S{=}197$ ($14{\times}14$ patches plus CLS token, 132 operator instances). All columns within a row share the same MicroBench primitive profile, so each comparison isolates placement; speedup is reported against the best static \emph{hybrid} baseline (BumbleBee/BOLT/BLB-style placements) and the \textbf{Mix} column gives the compiler's HE/MPC operator split (out of 132). Latencies use ``ks'' = $10^3\,$s for compactness.

\rev{The BERT-base tables additionally report the published BumbleBee system (BB$^{\mathrm{sys}}$) plotted in Figure~\ref{fig:sota_placement}(a,b) and the placement-fair speedup; the submitted version mistakenly repeated the ViT CPU table in place of the BERT-base CPU table. Table~\ref{tab:baselines-summary} summarizes the per-operator baselines over the full grid.}

\begin{table}[ht]
\centering\footnotesize\setlength{\tabcolsep}{3.5pt}
\caption{BERT-base, CPU, full $4\!\times\!4$ sweep. BB$^{\mathrm{sys}}$ is the published BumbleBee end-to-end system measurement recomposed under each network (the orange bars in Figure~\ref{fig:sota_placement}); BB$^{\mathrm{map}}$ applies BumbleBee's placement rule to our primitives. Speedup is against the fastest baseline including BB$^{\mathrm{sys}}$ (as plotted); Speedup$^{\mathrm{fair}}$ uses placement-fair mappings only.}
\label{tab:full-bert-cpu}
\begin{tabular}{c c c c c c c c c c}
\toprule
BW & RTT & BB$^{\mathrm{sys}}$ & BB$^{\mathrm{map}}$ & BOLT & BLB & Ours & Speedup & Speedup$^{\mathrm{fair}}$ & Mix \\
\midrule
10\,Mbps & 1\,ms & 10.8\,ks & 16\,ks & 14.4\,ks & 12.8\,ks & 7.93\,ks & $1.36\times$ & $1.62\times$ & 12/120 \\
10\,Mbps & 20\,ms & 22\,ks & 16\,ks & 14.4\,ks & 12.9\,ks & 7.95\,ks & $1.62\times$ & $1.62\times$ & 12/120 \\
10\,Mbps & 40\,ms & 33.7\,ks & 16.1\,ks & 14.5\,ks & 12.9\,ks & 7.97\,ks & $1.61\times$ & $1.61\times$ & 12/120 \\
10\,Mbps & 80\,ms & 57.2\,ks & 16.1\,ks & 14.5\,ks & 12.9\,ks & 8\,ks & $1.61\times$ & $1.61\times$ & 12/120 \\
\midrule
100\,Mbps & 1\,ms & 2.67\,ks & 9.18\,ks & 9.02\,ks & 8.86\,ks & 1.27\,ks & $2.11\times$ & $6.99\times$ & 0/132 \\
100\,Mbps & 20\,ms & 13.8\,ks & 9.19\,ks & 9.02\,ks & 8.86\,ks & 1.3\,ks & $6.84\times$ & $6.84\times$ & 0/132 \\
100\,Mbps & 40\,ms & 25.6\,ks & 9.19\,ks & 9.03\,ks & 8.87\,ks & 1.33\,ks & $6.69\times$ & $6.69\times$ & 0/132 \\
100\,Mbps & 80\,ms & 49\,ks & 9.2\,ks & 9.04\,ks & 8.88\,ks & 1.39\,ks & $6.41\times$ & $6.41\times$ & 0/132 \\
\midrule
1\,Gbps & 1\,ms & 1.86\,ks & 8.49\,ks & 8.48\,ks & 8.46\,ks & 457\,s & $4.06\times$ & $18.50\times$ & 0/132 \\
1\,Gbps & 20\,ms & 13\,ks & 8.5\,ks & 8.48\,ks & 8.47\,ks & 486\,s & $17.43\times$ & $17.43\times$ & 0/132 \\
1\,Gbps & 40\,ms & 24.7\,ks & 8.51\,ks & 8.49\,ks & 8.47\,ks & 516\,s & $16.42\times$ & $16.42\times$ & 0/132 \\
1\,Gbps & 80\,ms & 48.2\,ks & 8.52\,ks & 8.5\,ks & 8.48\,ks & 576\,s & $14.73\times$ & $14.73\times$ & 0/132 \\
\midrule
3\,Gbps & 1\,ms & 1.79\,ks & 8.44\,ks & 8.44\,ks & 8.43\,ks & 397\,s & $4.52\times$ & $21.22\times$ & 0/132 \\
3\,Gbps & 20\,ms & 12.9\,ks & 8.45\,ks & 8.44\,ks & 8.44\,ks & 426\,s & $19.81\times$ & $19.81\times$ & 0/132 \\
3\,Gbps & 40\,ms & 24.7\,ks & 8.45\,ks & 8.45\,ks & 8.44\,ks & 456\,s & $18.52\times$ & $18.52\times$ & 0/132 \\
3\,Gbps & 80\,ms & 48.2\,ks & 8.47\,ks & 8.46\,ks & 8.45\,ks & 516\,s & $16.38\times$ & $16.38\times$ & 0/132 \\
\bottomrule
\end{tabular}
\end{table}

\begin{table}[ht]
\centering\footnotesize\setlength{\tabcolsep}{3.5pt}
\caption{BERT-base, GPU, full $4\!\times\!4$ sweep. BB$^{\mathrm{sys}}$ is the published BumbleBee end-to-end system measurement recomposed under each network (the orange bars in Figure~\ref{fig:sota_placement}); BB$^{\mathrm{map}}$ applies BumbleBee's placement rule to our primitives. Speedup is against the fastest baseline including BB$^{\mathrm{sys}}$ (as plotted); Speedup$^{\mathrm{fair}}$ uses placement-fair mappings only.}
\label{tab:full-bert-gpu}
\begin{tabular}{c c c c c c c c c c}
\toprule
BW & RTT & BB$^{\mathrm{sys}}$ & BB$^{\mathrm{map}}$ & BOLT & BLB & Ours & Speedup & Speedup$^{\mathrm{fair}}$ & Mix \\
\midrule
10\,Mbps & 1\,ms & 10.8\,ks & 10.9\,ks & 9.34\,ks & 7.75\,ks & 472\,s & $16.41\times$ & $16.41\times$ & 132/0 \\
10\,Mbps & 20\,ms & 22\,ks & 11\,ks & 9.36\,ks & 7.76\,ks & 472\,s & $16.44\times$ & $16.44\times$ & 132/0 \\
10\,Mbps & 40\,ms & 33.7\,ks & 11\,ks & 9.37\,ks & 7.78\,ks & 472\,s & $16.47\times$ & $16.47\times$ & 132/0 \\
10\,Mbps & 80\,ms & 57.2\,ks & 11\,ks & 9.41\,ks & 7.81\,ks & 472\,s & $16.53\times$ & $16.53\times$ & 132/0 \\
\midrule
100\,Mbps & 1\,ms & 2.67\,ks & 1.46\,ks & 1.3\,ks & 1.14\,ks & 472\,s & $2.42\times$ & $2.42\times$ & 132/0 \\
100\,Mbps & 20\,ms & 13.8\,ks & 1.48\,ks & 1.32\,ks & 1.16\,ks & 472\,s & $2.45\times$ & $2.45\times$ & 132/0 \\
100\,Mbps & 40\,ms & 25.6\,ks & 1.49\,ks & 1.33\,ks & 1.17\,ks & 472\,s & $2.48\times$ & $2.48\times$ & 132/0 \\
100\,Mbps & 80\,ms & 49\,ks & 1.53\,ks & 1.36\,ks & 1.2\,ks & 472\,s & $2.55\times$ & $2.55\times$ & 132/0 \\
\midrule
1\,Gbps & 1\,ms & 1.86\,ks & 515\,s & 498\,s & 482\,s & 97.8\,s & $4.93\times$ & $4.93\times$ & 1/131 \\
1\,Gbps & 20\,ms & 13\,ks & 530\,s & 513\,s & 497\,s & 126\,s & $3.94\times$ & $3.94\times$ & 3/129 \\
1\,Gbps & 40\,ms & 24.7\,ks & 546\,s & 529\,s & 512\,s & 148\,s & $3.45\times$ & $3.45\times$ & 36/96 \\
1\,Gbps & 80\,ms & 48.2\,ks & 578\,s & 560\,s & 542\,s & 186\,s & $2.92\times$ & $2.92\times$ & 48/84 \\
\midrule
3\,Gbps & 1\,ms & 1.79\,ks & 444\,s & 439\,s & 433\,s & 38\,s & $11.39\times$ & $11.39\times$ & 0/132 \\
3\,Gbps & 20\,ms & 12.9\,ks & 460\,s & 454\,s & 448\,s & 66.7\,s & $6.71\times$ & $6.71\times$ & 1/131 \\
3\,Gbps & 40\,ms & 24.7\,ks & 476\,s & 469\,s & 463\,s & 90.5\,s & $5.11\times$ & $5.11\times$ & 48/84 \\
3\,Gbps & 80\,ms & 48.2\,ks & 508\,s & 500\,s & 493\,s & 127\,s & $3.87\times$ & $3.87\times$ & 48/84 \\
\bottomrule
\end{tabular}
\end{table}

\begin{table}[ht]
\centering\small
\caption{ViT-B/16, CPU, full $4\!\times\!4$ sweep. Bold marks the best RTT per bandwidth (the value shown in Figure~\ref{fig:sota_placement}).}
\label{tab:full-vit-cpu}
\begin{tabular}{c c c c c c c c}
\toprule
BW & RTT & BumbleBee & BOLT & BLB & Ours & Speedup & Mix HE/MPC \\
\midrule
10\,Mbps & 1\,ms & 28.2\,ks & 25.4\,ks & 22.6\,ks & 13\,ks & $\mathbf{1.75\times}$ & 12/120 \\
10\,Mbps & 20\,ms & 28.2\,ks & 25.4\,ks & 22.6\,ks & 13\,ks & $1.74\times$ & 12/120 \\
10\,Mbps & 40\,ms & 28.2\,ks & 25.4\,ks & 22.6\,ks & 13\,ks & $1.74\times$ & 12/120 \\
10\,Mbps & 80\,ms & 28.2\,ks & 25.4\,ks & 22.7\,ks & 13\,ks & $1.74\times$ & 12/120 \\
\midrule
100\,Mbps & 1\,ms & 15.9\,ks & 15.6\,ks & 15.3\,ks & 2.05\,ks & $\mathbf{7.48\times}$ & 0/132 \\
100\,Mbps & 20\,ms & 15.9\,ks & 15.6\,ks & 15.3\,ks & 2.08\,ks & $7.38\times$ & 0/132 \\
100\,Mbps & 40\,ms & 15.9\,ks & 15.6\,ks & 15.3\,ks & 2.11\,ks & $7.27\times$ & 0/132 \\
100\,Mbps & 80\,ms & 15.9\,ks & 15.6\,ks & 15.4\,ks & 2.17\,ks & $7.08\times$ & 0/132 \\
\midrule
1\,Gbps & 1\,ms & 14.7\,ks & 14.6\,ks & 14.6\,ks & 739\,s & $\mathbf{19.75\times}$ & 0/132 \\
1\,Gbps & 20\,ms & 14.7\,ks & 14.6\,ks & 14.6\,ks & 768\,s & $19.02\times$ & 0/132 \\
1\,Gbps & 40\,ms & 14.7\,ks & 14.6\,ks & 14.6\,ks & 798\,s & $18.32\times$ & 0/132 \\
1\,Gbps & 80\,ms & 14.7\,ks & 14.7\,ks & 14.6\,ks & 858\,s & $17.05\times$ & 0/132 \\
\midrule
3\,Gbps & 1\,ms & 14.6\,ks & 14.6\,ks & 14.5\,ks & 642\,s & $\mathbf{22.65\times}$ & 0/132 \\
3\,Gbps & 20\,ms & 14.6\,ks & 14.6\,ks & 14.6\,ks & 671\,s & $21.70\times$ & 0/132 \\
3\,Gbps & 40\,ms & 14.6\,ks & 14.6\,ks & 14.6\,ks & 701\,s & $20.78\times$ & 0/132 \\
3\,Gbps & 80\,ms & 14.6\,ks & 14.6\,ks & 14.6\,ks & 761\,s & $19.15\times$ & 0/132 \\
\bottomrule
\end{tabular}
\end{table}

\begin{table}[ht]
\centering\small
\caption{ViT-B/16, GPU, full $4\!\times\!4$ sweep. Bold marks the best RTT per bandwidth (the value shown in Figure~\ref{fig:sota_placement}).}
\label{tab:full-vit-gpu}
\begin{tabular}{c c c c c c c c}
\toprule
BW & RTT & BumbleBee & BOLT & BLB & Ours & Speedup & Mix HE/MPC \\
\midrule
10\,Mbps & 1\,ms & 19.1\,ks & 16.3\,ks & 13.5\,ks & 789\,s & $17.13\times$ & 132/0 \\
10\,Mbps & 20\,ms & 19.1\,ks & 16.3\,ks & 13.5\,ks & 789\,s & $17.15\times$ & 132/0 \\
10\,Mbps & 40\,ms & 19.1\,ks & 16.3\,ks & 13.5\,ks & 789\,s & $17.17\times$ & 132/0 \\
10\,Mbps & 80\,ms & 19.1\,ks & 16.3\,ks & 13.6\,ks & 789\,s & $\mathbf{17.21\times}$ & 132/0 \\
\midrule
100\,Mbps & 1\,ms & 2.54\,ks & 2.26\,ks & 1.99\,ks & 789\,s & $2.52\times$ & 132/0 \\
100\,Mbps & 20\,ms & 2.55\,ks & 2.28\,ks & 2\,ks & 789\,s & $2.54\times$ & 132/0 \\
100\,Mbps & 40\,ms & 2.57\,ks & 2.29\,ks & 2.01\,ks & 789\,s & $2.55\times$ & 132/0 \\
100\,Mbps & 80\,ms & 2.6\,ks & 2.32\,ks & 2.04\,ks & 789\,s & $\mathbf{2.59\times}$ & 132/0 \\
\midrule
1\,Gbps & 1\,ms & 888\,s & 860\,s & 832\,s & 161\,s & $\mathbf{5.17\times}$ & 1/131 \\
1\,Gbps & 20\,ms & 904\,s & 875\,s & 847\,s & 189\,s & $4.48\times$ & 3/129 \\
1\,Gbps & 40\,ms & 920\,s & 891\,s & 862\,s & 216\,s & $3.99\times$ & 36/96 \\
1\,Gbps & 80\,ms & 952\,s & 922\,s & 892\,s & 259\,s & $3.44\times$ & 36/96 \\
\midrule
3\,Gbps & 1\,ms & 766\,s & 757\,s & 747\,s & 62\,s & $\mathbf{12.05\times}$ & 0/132 \\
3\,Gbps & 20\,ms & 781\,s & 771\,s & 761\,s & 90.7\,s & $8.39\times$ & 0/132 \\
3\,Gbps & 40\,ms & 797\,s & 787\,s & 776\,s & 121\,s & $6.43\times$ & 3/129 \\
3\,Gbps & 80\,ms & 829\,s & 818\,s & 806\,s & 164\,s & $4.92\times$ & 48/84 \\
\bottomrule
\end{tabular}
\end{table}

\section{MicroBench Measurement Methodology}
\label{appx:measurement-methodology}

This appendix describes how \textsc{MicroBench} constructs the deployment-specific
cost profile used by the compiler. The purpose of \textsc{MicroBench} is to
instantiate operator-level costs for the target hardware and network setting,
rather than to benchmark a complete hybrid system for every possible deployment.

\paragraph{Cost entries.}
For each operator, conversion, and HE maintenance action, \textsc{MicroBench}
records three quantities: local compute time \(c^{\mathrm{loc}}\), communication
volume \(b\), and round count \(r\). The local compute time is measured on the
target CPU or GPU backend. The communication volume and round count are obtained
from the corresponding HE/MPC protocol implementation and message-size
accounting. Given a deployment setting
\(\mathcal{D}=(H,\mathrm{BW},\mathrm{RTT})\), the compiler evaluates each action as
\[
    c_{\mathcal{D}}
    =
    c^{\mathrm{loc}}_{H}
    +
    \frac{8b}{\mathrm{BW}}
    +
    r\cdot \mathrm{RTT}.
\]
This decomposition separates hardware-dependent local computation from
network-dependent communication cost, allowing the planner to re-evaluate the
same operator profile under different bandwidth and RTT settings.

\paragraph{Profiling protocol.}
We discard warm-up iterations and report the mean over timed iterations.
Table~\ref{tab:microbench-protocol} summarizes the profiling protocol used by
\textsc{MicroBench}. HE operators and conversions are measured with wall-clock
timing, while GPU MPC operators use CUDA-event timing to reduce timing overhead
for short GPU kernels.

\begin{table}[t]
\centering
\small
\caption{\textbf{MicroBench profiling protocol.}
We discard warm-up iterations and report the mean over timed iterations.}
\label{tab:microbench-protocol}
\begin{tabular}{lccc}
\toprule
Primitive family & Warmups & Repeats & Timing method \\
\midrule
HE operators & 1 & 3 & wall-clock \\
HE$\leftrightarrow$MPC conversions & 2 & 5 & wall-clock \\
MPC operators on CPU & 1 & 3 & wall-clock \\
MPC operators on GPU & 1 & 3 & CUDA event \\
\bottomrule
\end{tabular}
\end{table}

\begin{table}[ht]
\centering\small
\caption{Per-operator baselines over the $4\times4$ grid (min--max latency ratio to \textsc{CipherFlow}; all plans scored by the same evaluator). EncFormer-style placement keeps linear operators in FHE and runs Softmax/LayerNorm/GELU in MPC with complex-packed (half-payload) conversions; under our profiles it coincides with the BLB column.}
\label{tab:baselines-summary}
\begin{tabular}{llcccc}
\toprule
Model & Platform & Greedy & Random best & Random median & EncFormer-style \\
\midrule
BERT-base & CPU & 1.02--1.37$\times$ & 1.63--10.92$\times$ & 1.86--16.31$\times$ & 1.61--21.22$\times$ \\
BERT-base & GPU & 1.01--5.75$\times$ & 1.74--14.02$\times$ & 2.08--18.58$\times$ & 2.42--16.53$\times$ \\
ViT-B/16 & CPU & 1.02--1.43$\times$ & 1.64--11.34$\times$ & 1.92--17.02$\times$ & 1.74--22.65$\times$ \\
ViT-B/16 & GPU & 1.02--5.39$\times$ & 1.88--13.94$\times$ & 2.27--18.57$\times$ & 2.52--17.21$\times$ \\
\bottomrule
\end{tabular}
\end{table}

\paragraph{Shape-keyed reuse.}
MicroBench profiles operators by shape rather than by layer identity. Each entry
is keyed by
\[
(\mathrm{op\_type},\mathrm{input\_shape},\mathrm{output\_shape},\mathrm{backend}).
\]
Thus, repeated Transformer blocks reuse the same profile entry whenever they
share the same operator type and tensor shape. This reduces profiling overhead
and ensures placement-fair comparison: all static baselines and our compiler
consume the same cached cost entry for the same operator shape under the same
deployment profile.

\paragraph{Representative cost entries.}
Table~\ref{tab:microbench-entries} reports representative entries used by the
planner. These entries illustrate the cost asymmetry that motivates
hardware-aware placement. HE operators often have high local compute cost but no
MPC-style communication rounds, while MPC operators are locally cheaper but
introduce communication volume and round-trip latency. Cross-domain conversions
also contribute non-negligible cost, so the compiler must jointly optimize
operator placement and conversion placement.

\begin{table}[ht]
\centering
\small
\caption{\textbf{Representative MicroBench cost entries} (BERT-base, $S{=}128$; conversions for a $[1,128,768]$ tensor). Each entry reports local compute time \(c^{\mathrm{loc}}\), communication volume \(b\), and round count \(r\); these are the exact values consumed by the planner. $^\dagger$Derived from the CPU entries with published NEXUS-CUDA speedups (MatMul CPU/20$\times$, bootstrapping CPU/30$\times$), because profiling them on the GPU ran out of memory.}
\label{tab:microbench-entries}
\begin{tabular}{llccc}
\toprule
Operator / action & Domain / backend & \(c^{\mathrm{loc}}\) & \(b\) & \(r\) \\
\midrule
Attention\_QK & HE / CPU & 1.55e5 ms & 0 & 0 \\
Attention\_QK & MPC / CPU & 211 ms & 6.2 MiB & 1 \\
Softmax & HE / CPU & 5.98e4 ms & 0 & 0 \\
Softmax & MPC / CPU & 2.72e3 ms & 79.8 MiB & 11 \\
GeLU & MPC / CPU & 1.05e3 ms & 30.9 MiB & 4 \\
HE$\rightarrow$MPC & Conv. / CPU & 10.7 ms & 37.5 MiB & 1 \\
MPC$\rightarrow$HE & Conv. / CPU & 10.5 ms & 37.5 MiB & 1 \\
Bootstrap & HE / CPU & 5.16e4 ms & 0 & 0 \\
\midrule
Attention\_QK$^\dagger$ & HE / GPU & 7.5e3 ms & 0 & 0 \\
Attention\_QK & MPC / GPU & 11.7 ms & 19.1 MiB & 3 \\
Softmax & HE / GPU & 264 ms & 0 & 0 \\
Softmax & MPC / GPU & 64.7 ms & 106.2 MiB & 15 \\
GeLU & MPC / GPU & 13.5 ms & 22.2 MiB & 3 \\
HE$\rightarrow$MPC & Conv. / GPU & 1.38 ms & 37.5 MiB & 1 \\
MPC$\rightarrow$HE & Conv. / GPU & 1.27 ms & 37.5 MiB & 1 \\
Bootstrap$^\dagger$ & HE / GPU & 1.68e3 ms & 0 & 0 \\
\bottomrule
\end{tabular}
\end{table}

\section{Cost-Model Validation}
\label{app:validation}
\begin{figure}[ht]
\includegraphics[width=0.9\linewidth]{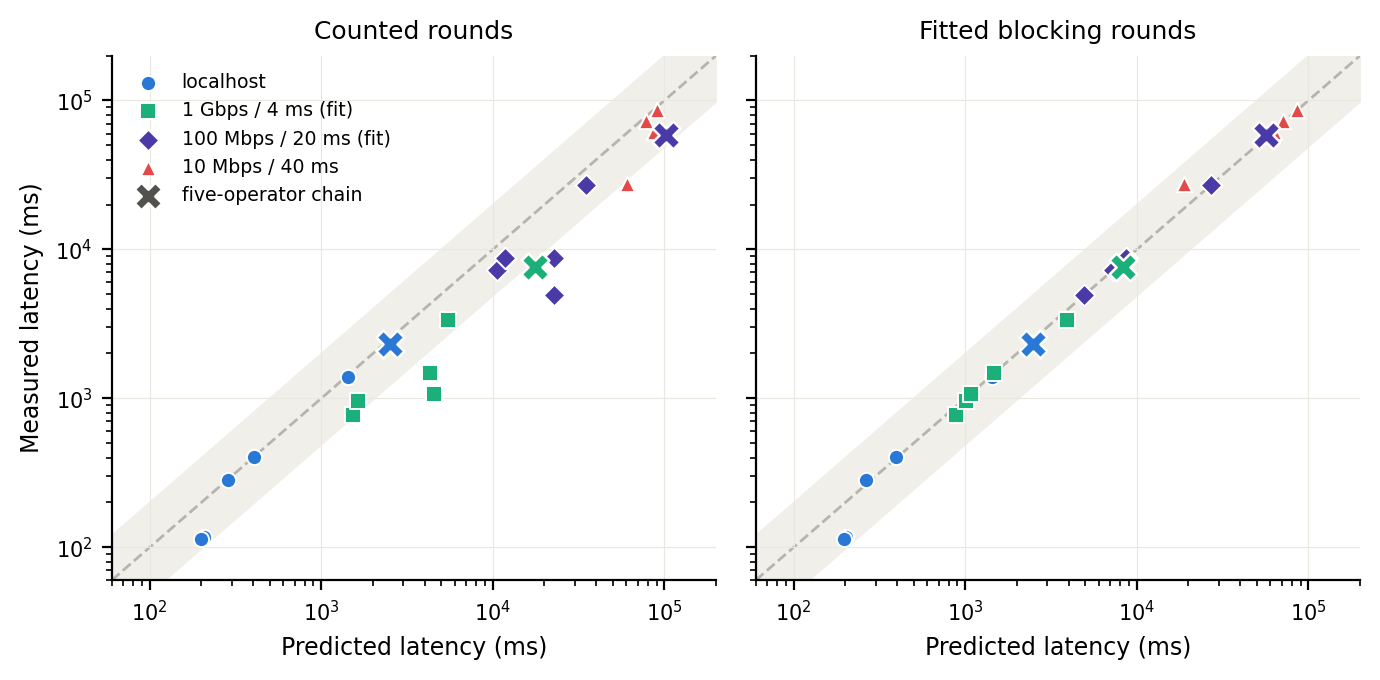}
\caption{Predicted versus measured latency of real SCI/BOLT protocol execution, per operator and for the five-operator chain (crosses), under localhost and \texttt{tc netem}-shaped networks, with counted rounds (left) and fitted blocking rounds (right). The fit uses the 1\,Gbps and 100\,Mbps per-operator runs; shading marks a factor of 2.}
\label{fig:validation}
\end{figure}

\rev{We executed the five nonlinear and attention operators of one encoder block ($B{=}1$, $S{=}16$) through the real SCI/BOLT bridges, both in isolation and back to back, and compared measured protocol wall-clock with the composed prediction $c^{\mathrm{loc}}+8b/\mathrm{BW}+r\cdot\mathrm{RTT}$. Networks were emulated with Linux \texttt{tc netem} on the loopback device. Figure~\ref{fig:validation} shows the result. The attention bridges run their 12 per-head products on 12 connections but count traffic on two, so their byte counts are scaled by 6. On localhost the chain is within 8--10\% of the prediction, with Softmax, LayerNorm, and GeLU within $\pm8\%$ and the two attention matrix multiplications 43--44\% faster than predicted. Under shaped networks the chain-level prediction is conservative (measured 52--58\% and 41--44\% below prediction at 1\,Gbps/4\,ms and 100\,Mbps/20\,ms). The error comes from the round term: the protocol round counter reports about 1{,}000 rounds for Softmax, but the implied blocking round trips are only about 175, because most rounds are pipelined; charging raw rounds overestimates round-heavy operators. We therefore refine the network term: for each operator, an effective volume $b_{\mathrm{eff}}$ and effective blocking rounds $r_{\mathrm{blk}}$ are fitted (non-negative least squares) to its 1\,Gbps and 100\,Mbps runs, and the operator is charged $c^{\mathrm{loc}}+8b_{\mathrm{eff}}/\mathrm{BW}+r_{\mathrm{blk}}\cdot\mathrm{RTT}$. The fit gives 175 and 132 blocking rounds for Softmax and LayerNorm (counted: 1{,}024 and 856), none for the transfer-bound QK, AV, and GeLU, and volumes within 10\% of the counted ones except Softmax (14.5 vs.\ 24.6\,MB). On runs not used in the fit, the chain error falls from 52--58\% to 1--11\% at 1\,Gbps and from 41--44\% to 1--7\% at 100\,Mbps; at 10\,Mbps/40\,ms the per-operator error is within 3\% for QK, AV, and LayerNorm, while Softmax is under-predicted (18.9 vs.\ 27.5\,s, one repetition).}

\section{Implementation Details}
\label{app:impl}
\rev{\textbf{HE backend.} HE operators call NEXUS CKKS kernels through Python adapters. Kernel shape and packing contracts (e.g., LayerNorm on $[B,S,768]$ without affine parameters, packed $[3,B,S,768]$ attention inputs with 12 heads) are checked explicitly; a violated contract marks the HE transition infeasible rather than silently approximating. \textbf{MPC backend.} MPC operators run in external SCI/BOLT or SHAFT processes for both parties; for SCI/BOLT, communication bytes and rounds are read from the protocol implementation's own counters, so $(b,r)$ are deterministic and only local compute is timed; for SHAFT, local compute is timed per operator and the bytes and rounds measured over a full SHAFT run are apportioned to operators by compute time. The SCI linear bridge caps its batch dimension, and larger shapes are tiled. \textbf{Conversions.} HE$\rightarrow$MPC uses mask-and-decrypt and MPC$\rightarrow$HE uses re-encryption; the directions are asymmetric and are profiled separately per tensor shape. \textbf{Profiling.} Warm-up iterations are discarded, short GPU kernels are timed with CUDA events, and profiles are cached by (operator, input shape, output shape, backend) so repeated blocks are profiled once.}

\section{Communication Volume}
\label{app:volume}
\begin{table}[ht]
\centering\small
\caption{Online HE--MPC protocol communication (GiB per BERT-base inference) of the latency-optimal plan versus static hybrid placements.}
\label{tab:comm-volume}
\begin{tabular}{lcccccc}
\toprule
Platform / network & Ours latency & Ours & BumbleBee & BOLT & BLB & All-MPC \\
\midrule
GPU, 10\,Mbps / 1\,ms & 472\,s & 0.0 & 12.3 & 10.4 & 8.5 & 10.5 \\
GPU, 3\,Gbps / 80\,ms & 127\,s & 9.4 & 12.3 & 10.4 & 8.5 & 10.5 \\
CPU, 10\,Mbps / 1\,ms & 7.93\,ks & 7.9 & 8.9 & 7.0 & 5.2 & 10.5 \\
CPU, 1\,Gbps / 20\,ms & 486\,s & 10.5 & 8.9 & 7.0 & 5.2 & 10.5 \\
\bottomrule
\end{tabular}
\end{table}

\clearpage
\section*{NeurIPS Paper Checklist}

The checklist is designed to encourage best practices for responsible machine learning research, addressing issues of reproducibility, transparency, research ethics, and societal impact. Do not remove the checklist: {\bf The papers not including the checklist will be desk rejected.} The checklist should follow the references and follow the (optional) supplemental material.  The checklist does NOT count towards the page
limit. 

Please read the checklist guidelines carefully for information on how to answer these questions. For each question in the checklist:
\begin{itemize}
    \item You should answer \answerYes{}, \answerNo{}, or \answerNA{}.
    \item \answerNA{} means either that the question is Not Applicable for that particular paper or the relevant information is Not Available.
    \item Please provide a short (1--2 sentence) justification right after your answer (even for \answerNA). 
\end{itemize}

{\bf The checklist answers are an integral part of your paper submission.} They are visible to the reviewers, area chairs, senior area chairs, and ethics reviewers. You will also be asked to include it (after eventual revisions) with the final version of your paper, and its final version will be published with the paper.

The reviewers of your paper will be asked to use the checklist as one of the factors in their evaluation. While \answerYes{} is generally preferable to \answerNo{}, it is perfectly acceptable to answer \answerNo{} provided a proper justification is given (e.g., error bars are not reported because it would be too computationally expensive'' or ``we were unable to find the license for the dataset we used''). In general, answering \answerNo{} or \answerNA{} is not grounds for rejection. While the questions are phrased in a binary way, we acknowledge that the true answer is often more nuanced, so please just use your best judgment and write a justification to elaborate. All supporting evidence can appear either in the main paper or the supplemental material, provided in appendix. If you answer \answerYes{} to a question, in the justification please point to the section(s) where related material for the question can be found.

IMPORTANT, please:
\begin{itemize}
    \item {\bf Delete this instruction block, but keep the section heading ``NeurIPS Paper Checklist"},
    \item  {\bf Keep the checklist subsection headings, questions/answers and guidelines below.}
    \item {\bf Do not modify the questions and only use the provided macros for your answers}.
\end{itemize}

\begin{enumerate}

\item {\bf Claims}
    \item[] Question: Do the main claims made in the abstract and introduction accurately reflect the paper's contributions and scope?
    \item[] Answer: \answerYes{} %
    \item[] Justification: The abstract and Section~\ref{sec:intro} state the four contributions (hardware-aware formulation, MicroBench, state-expanded planning, BlockFold) and the up-to $16.5\times$ / $22\times$ speedups; these are supported by Section~\ref{sec:problem-formulation}, Section~3, and the experiments in Section~4 and Appendix~\ref{appx:seqlen}.

\item {\bf Limitations}
    \item[] Question: Does the paper discuss the limitations of the work performed by the authors?
    \item[] Answer: \answerYes{} %
    \item[] Justification: The Discussion at the end of Section~4 explicitly notes that \textsc{CipherFlow} optimizes single-request execution under a fixed deployment profile and does not yet model multi-client concurrency, batching, queueing, or shared-resource contention.

\item {\bf Theory assumptions and proofs}
    \item[] Question: For each theoretical result, does the paper provide the full set of assumptions and a complete (and correct) proof?
    \item[] Answer: \answerYes{} %
    \item[] Justification: The threat model and problem formulation are stated in Section~\ref{sec:preliminaries}--\ref{sec:problem-formulation}; the BlockFold complexity result is derived in full in Appendix~\ref{app:blockfold-complexity}, and the state-transition rules used by the planner are listed in Appendix~\ref{app:state_transitions}.

    \item {\bf Experimental result reproducibility}
    \item[] Question: Does the paper fully disclose all the information needed to reproduce the main experimental results of the paper to the extent that it affects the main claims and/or conclusions of the paper (regardless of whether the code and data are provided or not)?
    \item[] Answer: \answerYes{} %
    \item[] Justification: Section~\ref{subsec:set} specifies models, sequence lengths, hardware, network sweep, and HE/MPC backends; the algorithmic skeleton is given in Appendix~\ref{appx:pseudocode}, and the MicroBench profiling protocol and representative cost entries are reported in Appendix~\ref{appx:measurement-methodology}.

\item {\bf Open access to data and code}
    \item[] Question: Does the paper provide open access to the data and code, with sufficient instructions to faithfully reproduce the main experimental results, as described in supplemental material?
    \item[] Answer: \answerYes{} %
    \item[] Justification: The code link is in \href{https://github.com/loopleap-lab/CipherFlow}{\texttt{https://github.com/loopleap-lab/CipherFlow}}.

\item {\bf Experimental setting/details}
    \item[] Question: Does the paper specify all the training and test details (e.g., data splits, hyperparameters, how they were chosen, type of optimizer) necessary to understand the results?
    \item[] Answer: \answerYes{} %
    \item[] Justification: Section~\ref{subsec:set} reports HE level budget, polynomial degree, sequence lengths, BW/RTT sweep, and the static baselines; we evaluate inference latency only (no training), so optimizer/data-split fields do not apply.

\item {\bf Experiment statistical significance}
    \item[] Question: Does the paper report error bars suitably and correctly defined or other appropriate information about the statistical significance of the experiments?
    \item[] Answer: \answerYes{} %
    \item[] Justification: Explicit error bars are omitted because per-operator HE/MPC latency is dominated by deterministic protocol-level compute and communication, and the run-to-run variance is much smaller than the order-of-magnitude speedups we report.

\item {\bf Experiments compute resources}
    \item[] Question: For each experiment, does the paper provide sufficient information on the computer resources (type of compute workers, memory, time of execution) needed to reproduce the experiments?
    \item[] Answer: \answerYes{} %
    \item[] Justification: Section~\ref{subsec:set} specifies the compute platforms (AMD Ryzen Threadripper PRO 3955WX 16-core CPU; NVIDIA B200 GPU); end-to-end inference latency is reported in Section~4.2 / Appendix~\ref{appx:seqlen}, and compilation time is reported in Table~\ref{tab:compile_ablation}.
    
\item {\bf Code of ethics}
    \item[] Question: Does the research conducted in the paper conform, in every respect, with the NeurIPS Code of Ethics \url{https://neurips.cc/public/EthicsGuidelines}?
    \item[] Answer: \answerYes{} %
    \item[] Justification: The work does not involve human subjects, personal data, or scraped content; it uses public model architectures and standard cryptographic primitives, and we have reviewed and conformed to the NeurIPS Code of Ethics.

\item {\bf Broader impacts}
    \item[] Question: Does the paper discuss both potential positive societal impacts and negative societal impacts of the work performed?
    \item[] Answer: \answerYes{} %
    \item[] Justification: Section~\ref{sec:intro} motivates the work by privacy regulation and recent data-leakage incidents; by lowering the latency cost of HE/MPC inference, \textsc{CipherFlow} makes privacy-preserving deployment more practical (positive impact). We do not foresee a direct negative pathway, since the compiler operates on user-supplied models and produces only protocol-level execution plans.
    
\item {\bf Safeguards}
    \item[] Question: Does the paper describe safeguards that have been put in place for responsible release of data or models that have a high risk for misuse (e.g., pre-trained language models, image generators, or scraped datasets)?
    \item[] Answer: \answerNA{} %
    \item[] Justification: We do not release any pretrained generative models, image generators, or scraped datasets; the released artifact is a compiler that produces secure-inference execution plans, which carries no high misuse risk.

\item {\bf Licenses for existing assets}
    \item[] Question: Are the creators or original owners of assets (e.g., code, data, models), used in the paper, properly credited and are the license and terms of use explicitly mentioned and properly respected?
    \item[] Answer: \answerYes{} %
    \item[] Justification: We cite the original works for all reused assets, including the HE/MPC backends (NEXUS-CKKS, SCI/SHAFT) in Section~\ref{subsec:set} and the static hybrid baselines (BumbleBee, BOLT, BLB) in Section~\ref{sec:intro} and Section~4. All assets are used within the terms of their original licenses.

\item {\bf New assets}
    \item[] Question: Are new assets introduced in the paper well documented and is the documentation provided alongside the assets?
    \item[] Answer: \answerYes{} %
    \item[] Justification: The new asset (the \textsc{CipherFlow} compiler) is documented in the paper itself: Section~3 describes the compilation pipeline, Appendix~\ref{appx:pseudocode} provides the algorithmic skeleton, Appendix~\ref{app:state_transitions} lists the state-transition rules, and Appendix~\ref{appx:measurement-methodology} documents the MicroBench profiling methodology and representative cost entries.

\item {\bf Crowdsourcing and research with human subjects}
    \item[] Question: For crowdsourcing experiments and research with human subjects, does the paper include the full text of instructions given to participants and screenshots, if applicable, as well as details about compensation (if any)?
    \item[] Answer: \answerNA{} %
    \item[] Justification: The work does not involve crowdsourcing or research with human subjects.

\item {\bf Institutional review board (IRB) approvals or equivalent for research with human subjects}
    \item[] Question: Does the paper describe potential risks incurred by study participants, whether such risks were disclosed to the subjects, and whether Institutional Review Board (IRB) approvals (or an equivalent approval/review based on the requirements of your country or institution) were obtained?
    \item[] Answer: \answerNA{} %
    \item[] Justification: The work does not involve human subjects, so IRB approval is not required.

\item {\bf Declaration of LLM usage}
    \item[] Question: Does the paper describe the usage of LLMs if it is an important, original, or non-standard component of the core methods in this research? Note that if the LLM is used only for writing, editing, or formatting purposes and does \emph{not} impact the core methodology, scientific rigor, or originality of the research, declaration is not required.
    \item[] Answer: \answerNA{} %
    \item[] Justification: BERT and ViT are evaluated only as fixed inference workloads for the compiler; LLMs are not part of the core methodology, and any incidental use during writing did not affect scientific content.

\end{enumerate}

\end{document}